\documentclass[%
 reprint,
 amsmath,amssymb,
 aps,
]{revtex4-2}

\usepackage{graphicx}
\usepackage{dcolumn}
\usepackage{bm}
\usepackage{xcolor}
\usepackage{tabularx}

\begin{document}

\preprint{APS/123-QED}

\title{Piezoelectric optomechanical approaches for efficient quantum microwave-to-optical signal transduction: the need for co-design}

\author{Krishna C. Balram}
 \email{krishna.coimbatorebalram@bristol.ac.uk}
 \affiliation{Quantum Engineering Technology Labs and Department of Electrical and Electronic Engineering, University of Bristol, Woodland Road, Bristol, UK BS8 1UB}
 
\author{Kartik Srinivasan}%
 \email{kartik.srinivasan@nist.gov}
\affiliation{Microsystems and Nanotechnology Division, National Institute of Standards and Technology, Gaithersburg, MD 20899, USA}%
\affiliation{Joint Quantum Institute, NIST/University of Maryland, College Park, MD 20742, USA}%

\date{\today}

\begin{abstract}
Piezoelectric optomechanical platforms represent one of the most promising routes towards achieving quantum transduction of photons between the microwave and optical frequency domains. However, there are significant challenges to achieving near-unity transduction efficiency. We discuss such factors in the context of the two main approaches being pursued for high efficiency transduction. The first approach uses one-dimensional nanobeam optomechanical crystals excited by interdigitated transducers, and is characterized by large single-photon optomechanical coupling strength, limited intracavity pump photon population to avoid absorption-induced heating, and low phonon injection efficiency from the transducer to the optomechanical cavity. The second approach uses (quasi) bulk acoustic wave resonators integrated into photonic Fabry-Perot cavity geometries, and is characterized by low single-photon optomechanical coupling strength, high intracavity pump photon population without significant heating, and high phonon injection efficiency. After reviewing the current status of both approaches, we discuss the need for co-designing the electromechanical and optomechanical sub-systems in order to achieve high transduction efficiencies, taking the GaAs piezo-optomechanical platform as an example.
\end{abstract}

\maketitle

\section{Introduction}

As quantum computing platforms continue to mature, it is being increasingly recognized that every physical system, whether it be trapped ions \cite{monroe2013scaling}, superconducting qubits, electron spins \cite{bardin2021microwaves}, or large-scale integrated photonics~\cite{wang_multidimensional_2018}, has fundamental limitations and a hybrid systems approach \cite{kurizki2015quantum} combining the desirable properties of multiple physical systems may ultimately be necessary for implementing large-scale error-corrected quantum computers and highly functional repeater-based quantum networks. Such hybrid platforms by necessity require quantum transducers that can provide efficient quantum interfaces between the different physical platforms, which may operate at very different frequencies and in different physical environments. A canonical problem here is the transduction of quantum signals between the microwave and optical frequency domains~\cite{regal_cavity_2011,safavi-naeini_proposal_2011,lauk2020perspectives,chu2020perspective}. This task captures the interface problems originating with linking superconducting qubits and electron spin systems \cite{bardin2021microwaves}, which have transition frequencies in the 1 GHz to 10 GHz frequency range, and telecom optical frequencies (near 194 THz) for which the lowest loss routing in optical fibers is achieved. Such signal transducers are essential for building distributed quantum networks \cite{kimble2008quantum} based on superconducting qubit circuits as nodes and adding quantum memories (based on spin systems) to these networks. 

While efficient quantum frequency conversion has been demonstrated between photons in the optical frequency domain using nonlinear guided wave interactions \cite{raymer2012manipulating}, these ideas cannot be directly applied to the microwave-to-optical problem because of the disparity in the frequencies (and corresponding wavelengths). To give some perspective, the free space wavelength of a 3 GHz microwave photon is 10 cm, compared to 1.55 ${\mu}m$ for the telecom-band optical photon. Given that the interaction strength between the two fields scales with their overlap, this size disparity makes it challenging to achieve strong Kerr-type nonlinearities for photon conversion, although there has been some exciting progress on this electro-optic front recently, mainly driven by the use of superconducting RF cavities \cite{rueda_efficient_2016,fan_superconducting_2018,hease2020bidirectional,fu_ground-state_2020}. Piezoelectric optomechanical (or piezo-optomechanical) approaches~\cite{bochmann_nanomechanical_2013,balram_coherent_2016,jiang_efficient_2020,han_cavity_2020,mirhosseini2020superconducting} partially circumvent this size mismatch problem by converting the microwave signal into a mechanical mode that can now have the same wavelength as the optical signal, on account of the much slower acoustic wave velocity. The same 3 GHz RF signal, when converted to a 3 GHz sound wave, will have an acoustic wavelength of 1 ${\mu}m$, assuming a sound wave velocity of 3000 m/s. This allows strong acousto-optic interactions to be engineered in small mode volume dielectric optomechanical cavities, and is one reason for strong interest in the field of (quantum) cavity optomechanics \cite{aspelmeyer_cavity_2014,safavi2019controlling}, with the promise of observing and controlling quantum effects in (macroscopic) mechanical objects. Moreover, mechanical systems have been strongly coupled to qubits \cite{o2010quantum,satzinger2018quantum,chu2017quantum,moores2018cavity} and can exhibit ultra-long coherence times at GHz frequencies \cite{maccabe2020nano,gokhale2020epitaxial,kervinen2018interfacing}, suggesting that mechanically-mediated transduction can be complemented by a number of other important functions that can be realized in the mechanical domain.   

While significant progress has been made in using piezoelectric optomechanical devices for microwave-to-optical transduction (hereafter referred to as MW-OT), including operation with the mechanical resonator near its quantum ground state~\cite{forsch_microwave_2018}, acoustic wave engineering to better link piezoelectric transducers to optomechanical cavities~\cite{jiang_efficient_2020}, incorporating a microwave cavity to resonantly enhance the electromechanical interaction~\cite{han_cavity_2020},  and the demonstration of upconversion of superconducting qubit photons to the optical domain~\cite{mirhosseini2020superconducting}, the overall transduction efficiency (i.e., including all input/output coupling losses) has remained less than 1~$\%$.  In this perspective, we use two common device architectures as illustrative examples to outline the difficulties in realizing high overall transduction efficiency, which stems from the challenge of simultaneously realizing high piezolectric and optomechanical transduction efficiencies.  These architectures take as a starting point an optimized piezoelectric or optomechanical transducer geometry, and much effort has gone into trying to develop the complementary optomechanical or piezoelectric piece needed for a full microwave-to-optical transducer. After presenting the aforementioned challenges in successfully building full transducers based on these architectures, we will discuss how co-design of the piezoelectric and optomechanical pieces from the onset may lead to new advantageous device geometries that are specifically tailored for this demanding application.


\begin{figure*}

\includegraphics[width=0.7\linewidth]{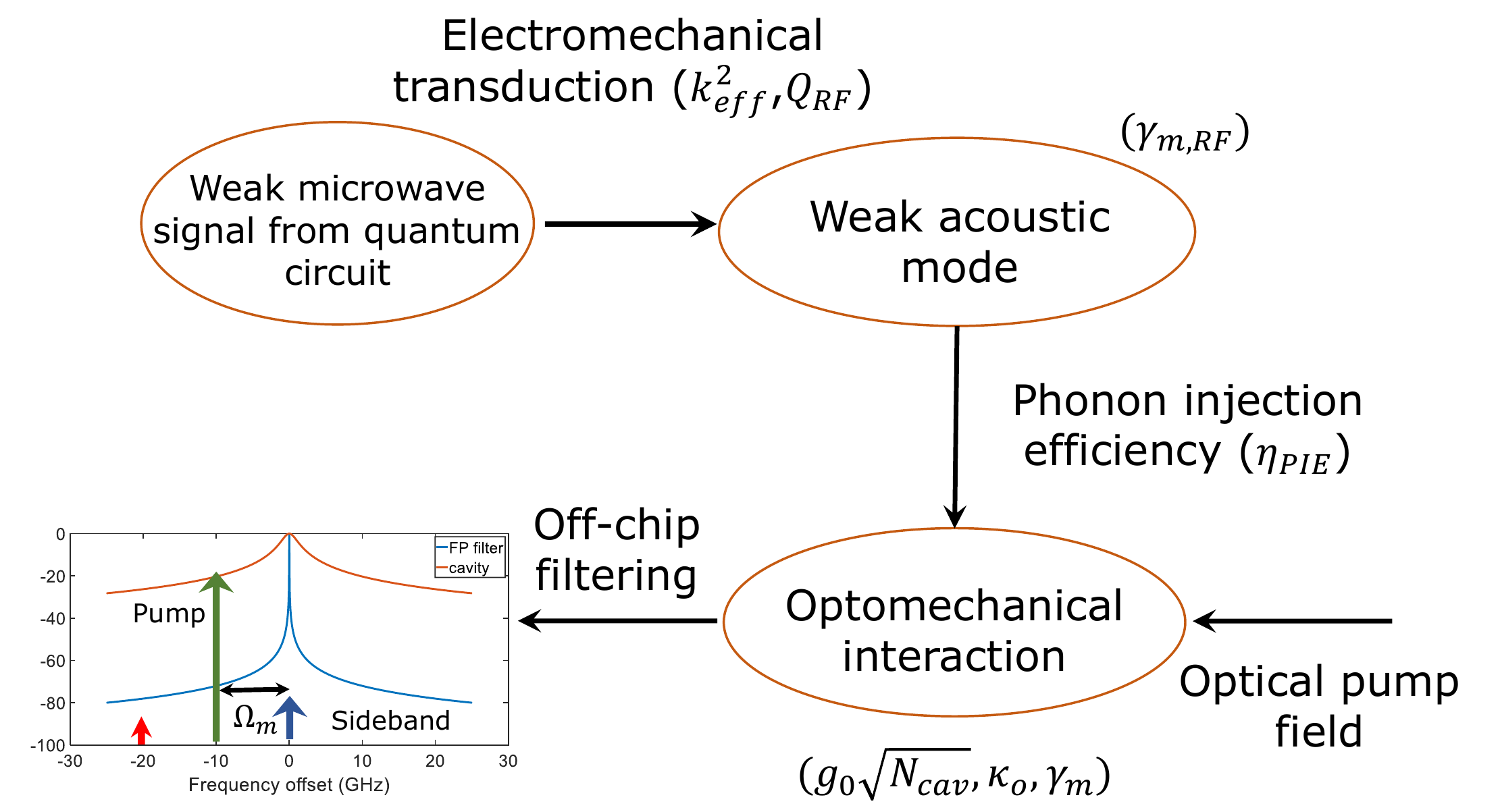}

\caption{\textbf{Schematic illustration of microwave-to-optical signal transduction using a piezo-optomechanical device.} The weak microwave signal (e.g., from a qubit) is first converted to an acoustic mode using a piezoelectric transducer. This acoustic mode is either co-located with an optical cavity (i.e., to form an optomechanical cavity) or else injected into a separate optomechanical cavity, where the parametric interaction mediated by a strong optical pump field up-converts the acoustic signal into an optical sideband. Though the optical cavity naturally provides some spectral filtering, external filtering will typically be needed to more fully suppress the optical pump and access the quantum state in the optical domain. Each element in the transduction process has certain key parameters that govern the efficiency. The piezoelectric transducer requires impedance matching and a high piezoelectric coefficient $k_{eff}^2$ to achieve efficient microwave-to-acoustic mode transduction. If the piezoelectric transducer and the optomechanical cavity are physically distinct, then the acoustic mode needs to be injected into the optomechanical cavity ($\eta_{PIE}$) with high efficiency. The optomechanical interaction is governed by the cooperativity $C_{om} = \frac{4g_{0}^2N_{cav}}{\gamma_{m}\kappa_{o}}$, where $g_{0}$ is the single-photon coupling rate, $N_{cav}$ is the number of intracavity pump photonics, and $\kappa$ and $\gamma$ are the optical and mechanical decay rates, respectively. As the strong optical pump needs to be externally filtered using fiber cavities, high mechanical mode frequencies which provide larger pump sideband frequency separation are desirable.}
\label{fig:Schematic}
\end{figure*}

\section{General considerations for piezo-optomechanical approaches}

Figure~\ref{fig:Schematic} presents a schematic illustration of MW-OT in piezo-optomechanical platforms. The conversion process can be thought of as being comprised of multiple stages \cite{chu2020perspective,lauk2020perspectives}. First, the microwave signal is coupled into the piezoelectric transducer, which is not necessarily trivial if the transducer has a high impedance and the microwave signal is carried by a standard 50~$\Omega$ transmission line. The piezoelectric transducer generates an acoustic phonon, which can be a propagating wave or a standing wave depending on the transducer geometry. This acoustic phonon is coupled into an opotomechanical cavity, which is used to realize upconversion into the optical domain. The optomechanical resonator realizes strong acousto-optic interaction through resonant enhancement (in both the optical and mechanical domains) of photoelastic and moving boundary effects, with the upconverted optical tone appearing as a sideband of the optical pump which drives the system (and which provides the photon energy needed to bridge the $\approx$200~THz difference between the input microwave and output optical frequencies). 

Such considerations are captured in the standard equation used to describe the efficiency of microwave-to-optical conversion in mechanically mediated schemes~\cite{ safavi-naeini_proposal_2011, Tian_adiabatic-state-conversion_2012,Wang_quantum-state-transfer_2012,Andrews2014,wu2020microwave}, given as: 

\begin{align}
\eta_\text{peak}  & =  \eta_\text{e} \eta_\text{o} \frac{4 \mathcal{C}_\text{EM} \mathcal{C}_\text{OM} \mathcal{L}^2_+}{(1 + \mathcal{C}_\text{EM} + \mathcal{C}_\text{OM} (\mathcal{L}^2_+ - \mathcal{L}^2_-))^2},
\label{eq:peakEta}
\end{align}

\noindent where $\eta_{e}$ is the microwave in-coupling efficiency, $\eta_{o}$ is the optical out-coupling efficiency, $C_\text{EM}$ and $C_\text{OM}$ are the electromechanical and optomechanical cooperativities, respectively, and $\mathcal{L}_\pm$ are the optical-cavity Lorentzian sideband amplitudes (see Fig.~\ref{fig:Schematic}), which essentially describe the amount of sideband resolution in the system, and are given by:

\begin{equation}
\mathcal{L}_\pm^2 = \frac{(\kappa_\text{o}/2)^2}{(\kappa_\text{o}/2)^2 + (\Delta \pm \Omega_\text{m})^2}
\label{eq:Lorentzian}
\end{equation}

where $\Delta$ represents the detuning of the optical pump from the optical cavity frequency. The electromechanical ($C_\text{EM}$) and optomechanical cooperativities ($C_\text{OM}$) are given by: 

\begin{equation}
\mathcal{C}_\text{EM} = \frac{4g^2_\text{EM}}{\gamma_\text{m} \kappa_\text{e}}, 
\label{eq:C-EM-def}
\end{equation}

\begin{eqnarray}
\mathcal{C}_\text{OM} = \frac{4g^2_\text{0}{N_{cav}}}{\gamma_\text{m} \kappa_\text{o}},
\label{eq:Com}
\end{eqnarray}

\noindent where $g_\text{EM}$ is the electromechanical coupling rate and is proportional to the piezoelectric coupling coefficient, $\sqrt{k_\text{eff}^2}$ (the latter more commonly used in the MEMS community \cite{hashimoto2009rf}), $\gamma_m$ is the decay rate of the mechanical mode, $\kappa_{o}$ and $\kappa_{e}$ are the decay rates of the optical and electrical modes, and $g_\text{0}$ is the optomechanical coupling rate. The mechanical ($Q_{m}$) and optical ($Q_{o}$) quality factors are defined as $Q_{m}=\Omega_{m}/\gamma_{m}$ and $Q_{o}=\omega_{o}/\kappa_{o}$ where $\Omega_{m}$ and $\omega_{o}$ are the mechanical and optical frequencies of the optomechanical cavity. $N_{cav}$ is the number of optical pump photons in the cavity.

As discussed above, the coupling between the optical and mechanical modes in an optomechanical cavity can be engineered using either the moving boundary (MB) or the photoelastic effect (PE) \cite{Balram_Moving_2014}. To compare and contrast these two effects, it is instructive to look at the optomechanical interaction in an idealized 1D Fabry-Perot cavity of length $L$. The resonant wavelength for the fundamental cavity mode can be approximated as $\lambda_{c} \sim 2nL$, where $n$ is the refractive index of the cavity medium. To first order, the change in the optical cavity wavelength ($\delta\lambda_{c}$) can be written as $\delta\lambda_{c}= n{\Delta}L+L{\Delta}n$. MB effects originate due to changes in the cavity dimensions ($\Delta L$) and PE effects due to local stress induced changes in the refractive index ($\Delta n$). Both effects are usually present in an optomechanical cavity, although their relative magnitudes and phases differ depending on the geometry in consideration. Usually, at high frequencies (e.g., in the GHz), where displacement amplitudes are low, the optomechanical coupling strength is primarily dominated by the PE effect. This tends to hold true, for example, in the localized optomechanical interactions in engineered 1D nanobeam optomechanical crystals and the travelling wave interactions seen in stimulated Brillouin scattering \cite{van2016unifying}.

For the 1D case, one can write down approximate expressions for the MB and PE effects. The RMS amplitude of the zero-point fluctuations of the mechanical mode at frequency $\Omega_m$ are given by $x_{zpf}=\sqrt{\hbar/2m_{eff}\Omega_{m}}$, where the effective mass of a 1D mechanical mode can be written in terms of material density ($\rho$), cavity cross-section ($A_\text{eff}$), and cavity length ($L$) as $m_{eff}={\rho}A_{eff}L$. The MB contribution can then be approximated as $g_{0}=g_{om}x_{zpf}$ where $g_{om}=d\omega/dL$ for a 1D Fabry-Perot cavity is given by $g_{om}=\omega_{c}/L$. This leads to:

\begin{eqnarray}
g_\text{0,MB} \sim \frac{\omega_{c}}{L^{3/2}}\sqrt{\frac{\hbar}{2{\rho}A_{eff}\Omega_{m}}}
\label{eq:g0_mb}
\end{eqnarray}

\noindent Similarly, the photoelastic contribution when the optical and mechanical modes are phase matched (the Brillouin scattering condition) is given by \cite{renninger2018bulk,valle_high-frequency_2019}:

\begin{eqnarray}
g_\text{0,PE} \sim \frac{\omega_{c}^2n^3p_{12}}{2c}\sqrt{\frac{\hbar}{2{\rho}A_{eff}L\Omega_{m}}}
\label{eq:g0_mb}
\end{eqnarray}

\noindent where $p_{12}$ is the photoelastic coefficient mediating the interaction between the optical and mechanical modes and $c$ is the speed of light. 

Despite the very different physical origins of the optomechanical interaction, both $g_{0,\text{MB}}$ and $g_{0,\text{PE}}$ can be significantly enhanced by moving to small mode volume (small $L$ and small $A_\text{eff}$) optomechanical cavities, a recurring theme in this work. It is important to note that current state of the art superconducting qubits have transition frequencies in the 3 GHz to 10 GHz frequency range and as discussed above, the PE dominates in that regime. High mechanical frequencies are also beneficial from the perspective of optical filtering, as discussed later in this article. To ensure strong optomechanical interactions using PE, a general rule of thumb is to engineer the interactions in small mode volume cavities in materials with large refractive indices ($n$) and photoelastic coefficients ($p_{12}$). This approach has been applied to realize large $g_{0}$ ($\approx$ 1 MHz) in 1D optomechanical crystal cavities in semiconductors like silicon and GaAs.

In addition to conversion efficiency, one needs to minimize the added noise during the conversion process to preserve the fidelity of the quantum state after transduction. The noise in MW-OT originates from two fundamental sources.  One is due to the thermal population of the mechanical resonator, and is given by: 

\begin{eqnarray}
N_\text{m} = \frac{1}{\eta_\text{e}} \frac{n_\text{m}}{\mathcal{C}_\text{EM}}, 
\label{eq:Nm}
\end{eqnarray}

\noindent where $n_\text{m}(\omega)=(e^{\hbar\omega/(k_{\text{B}}T)}-1)^{-1}$ is the thermal occupation of the mechanical resonator due to its contact with a thermal bath at temperature $T$, and with $k_\text{B}$ and $\hbar$ being the Boltzmann constant and the Planck constant divided by 2$\pi$, respectively.  

The other noise source is Stokes scattering of the optical drive field off the mechanical resonator, that is, the red-detuned sideband shown in Fig.~\ref{fig:Schematic}.  It is given by:

\begin{equation}
N_\text{o} = \frac{1}{\eta_\text{e}} \frac{\mathcal{C}_\text{OM}\mathcal{L}^2_-}{\mathcal{C}_\text{EM}},
\label{eq:Nopt1}
\end{equation}

These equations provide a generic prescription for simultaneously realizing high transduction efficiency and low noise. In particular, low noise is achieved by limiting the thermal population of the mechanical resonator, most commonly by operating in the quantum ground state via cryogenic cooling of a high frequency mechanical mode, and working in the sideband-resolved regime so that $\mathcal{L}_{-}$ is small (which also benefits from a high frequency mechanical mode). High transduction efficiency is achieved for large $C_\text{EM}$ and $C_\text{OM}$ (ideally $C_\text{EM}$ = $C_\text{OM}$+1 in the limit of small $\mathcal{L}_{-}$); physically, this essentially means that the system simultaneously achieves high conversion efficiency from a microwave signal to the acoustic mode of interest (that, the acoustic mode that is well-coupled to an optical mode) and efficient scattering of optical photons by that acoustic mode. These requirements are quite challenging to simultaneously realize because ultimately, the same mechanical mode must be linked to the microwave domain and the optical domain. This means that large $C_\text{OM}$ and $C_\text{EM}$ must be achieved for geometries in which the mechanical mode can be both piezoelectrically and optomechanically accessed while separating the superconducting metals to be used in the former from the photonic cavities and waveguides to be used in the latter. 

\begin{figure*}
\begin{center}
\includegraphics[width=0.8\linewidth]{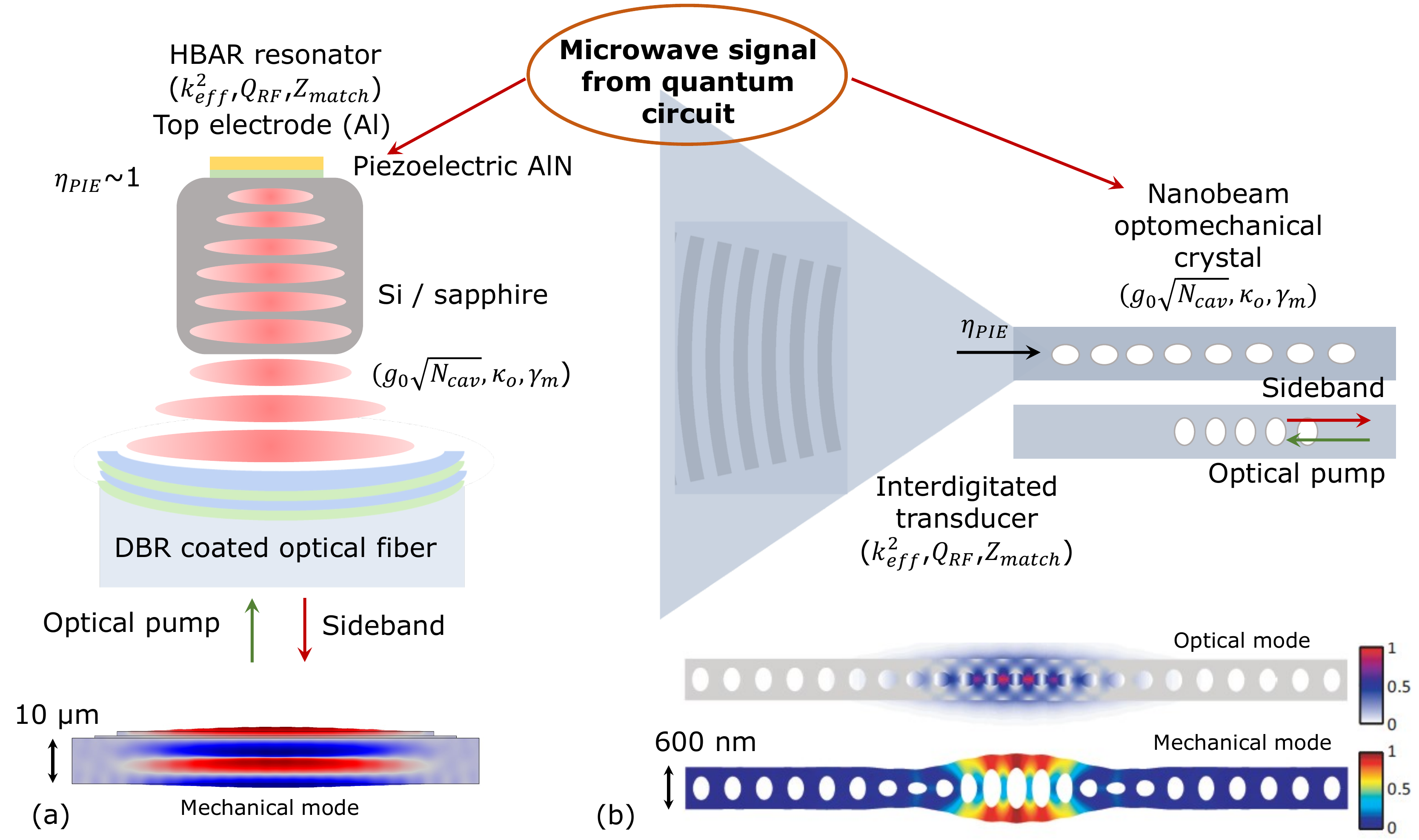}
\end{center}
\caption{Two complementary approaches towards piezoelectric microwave-to-optical signal transduction. (a) Optomechanical interaction in a high overtone bulk acoustic resonator built inside small mode volume fiber Fabry-Perot cavity (pictorial illustration of the optical field superimposed over the resonator schematic; mechanical mode simulation shown at the bottom). While the $\eta_{PIE}$ can approach 1 in these geometries and they are ideal from the microwave side, the small $g_{0}$, due to the large cavity mode volume, requires compensation by a large intracavity photon number ($N_{cav}$). (b) 1D nanobeam optomechanical crystal with high $g_{0}$ is used for the optomechanical interaction (optical and mechanical mode simulations shown at the bottom). The mechanical mode of the cavity is effectively electrically driven by an interdigitated transducer (IDT). The key challenge here lies in engineering sufficiently high $\eta_{PIE}$ to link the IDT-driven acoustic wave to the nanobeam breathing mode, accounting for mode size mismatch. Note the very different size scales of the two approaches.}
\label{fig:Approaches}
\end{figure*}


To address the challenge associated with simultaneously realizing low microwave losses and low optical losses in a system with a shared mechanical resonance, researchers have considered the use of interdigitated transducer (IDT) piezoelectric elements~\cite{campbell_surface_1998} that create a propagating acoustic wave from a microwave input. This propagating acoustic wave can then couple to the optomechanical cavity~\cite{balram_coherent_2016,Vainsencher_Bi_2016,jiang_efficient_2020}, with the length of acoustic waveguide providing effective spatial separation between the microwave elements and the optical elements. In such schemes, the peak microwave-to-optical transduction efficiency from Eq.~(\ref{eq:peakEta}) needs to be modified by a multiplicative factor ${\eta}_{PIE}$, which represents the phonon injection efficiency. This parameter essentially describes how well the piezoelectrically-excited acoustic wave couples to the mechanical cavity. In practice, ${\eta}_{PIE}$ is a critical parameter because, at a certain level, the piezoelectric MEMS community~\cite{gong_design_2013} and optomechanics community~\cite{chan_optimized_2012} have already individually developed electromechanical and optomechanical systems achieving high cooperativities, and the extent to which these known paradigms can be combined without degradation is one of the main underlying challenges for MW-OT. 

To engineer effective MW-OTs, one might chose to build upon existing device architectures from two complementary perspectives. We can either take existing devices that have high $C_\text{EM}$~\cite{hashimoto2009rf,gokhale2020epitaxial} and build high optical $Q_{o}$, small mode volume cavities~\cite{hunger2010fiber} around them in such a way that high $C_\text{OM}$ is realized and ${\eta}_{PIE}$ is naturally near-unity (i.e., no mechanical mode conversion), or we can start with devices that have the strongest acousto-optic interaction and therefore a high $C_\text{OM}$~\cite{chan_optimized_2012,Balram_Moving_2014} and understand how to connect these devices to a piezoelectric transducer that can achieve high $C_\text{EM}$ and high ${\eta}_{PIE}$. A schematic overview of these approaches is shown in Fig.~\ref{fig:Approaches}.  

The first approach is represented by bulk acoustic wave resonators combined with optical cavities \cite{chu2017quantum, kervinen2018interfacing, valle_high-frequency_2019}, while the second approach utilizes different piezoelectric actuation mechanisms to acoustically couple to a one-dimensional photonic/phononic crystal nanobeam (often termed an optomechanical crystal) \cite{balram_coherent_2016, bochmann_nanomechanical_2013, jiang_lithium_2019,mirhosseini2020superconducting}. While significant progress has been made on both fronts, they each have basic limitations that apparently constrain the overall achievable transduction efficiency. The physical origins of these limitations will be discussed in the next two sections. 

\section{Transduction using high-${\eta}_{PIE}$ bulk acoustic wave resonators}

We first start by considering microwave-to-optical transduction using optical cavities built around high overtone bulk acoustic wave resonators (HBARs). In the 2~GHz to 8~GHz frequency range, bulk acoustic wave devices (illustrated in Fig.~\ref{fig:Approaches}(a)), which lie at the heart of filtering in modern cell phones, have traditionally shown the highest microwave-to-acoustic wave conversion efficiency \cite{hashimoto2009rf}, with power injection efficiency from transducer to phonon cavity approaching unity~\cite{gokhale2020epitaxial}. In addition to efficient conversion of microwaves into acoustic waves, the phonons are excited predominantly in the mode of interest (large $\eta_{PIE}$), with very little spurious mode excitation, which can be controlled by modifying the transducer shape. While these resonators are almost ideal from a microwave-to-phonon conversion perspective, their size (surface area) is usually in the 10~${\mu}m^2$ to 100~${\mu}m^2$ range and the resonator geometry usually restricts the optical cavity architectures to simple Fabry-Perot variants with large mode volumes ($\gtrsim$ 5${\lambda}^3$ to 10${\lambda}^3$). As discussed above, engineering strong optomechanical interactions requires small mode volume optical cavities, which is more challenging to do with these bulk devices.

The challenge with effectively building optical cavities around HBAR resonators is not just a question of cavity mode volume, but also optical quality factor $Q_{o}$.  This challenge can be understood from the design of the HBAR devices, which confine the acoustic field in the thickness modes predominantly under the metal electrode. This forces one of the mirrors in the optical cavity to be made of metal, ultimately limiting the achievable $Q_{o}$ (metallic film reflectivities are $\leq$ 0.985 in the telecom band). In principle, one can avoid using the metal electrode for the HBARs, by using transparent electrodes made with graphene or Indium tin oxide (ITO), but these approaches come with the inherent tradeoff of higher series resistance, which lowers the $\eta_{PIE}$. As an alternative way of avoiding the metal mirror, one can always choose to work with shear modes which can be confined between metals (with the optical field in-between, thereby avoiding overlap with the metal), but the tradeoff then becomes a lower electromechanical coupling coefficient and lower mechanical quality factor $Q_{m}$, leading to an overall reduced $\mathcal{C}_\text{EM}$. These challenges effectively point to the need for co-design approaches, where we can ask the question, what is the best geometry for buiding small mode volume, high $Q_{o}$ optical cavities around HBAR resonators while still preserving their main advantages, high $Q_{m}$ and high $\eta_{PIE}$?

In addition to the optical cavity design problems, there are some additional constraints imposed on the HBAR geometry by the qubit frequencies. Most current superconducting qubit implementations work in the 3-10 GHz frequency range. It is challenging to engineer strong modal overlap between the optical and mechanical fields at these frequencies in a standard HBAR geometry, which restricts the achievable $g_{0}$ in these geometries, especially compared to 1D nanobeam optomechanical crystals. For example, if we take a simple high overtone bulk acoustic wave resonator in silicon and operate the device collinearly \cite{valle_high-frequency_2019}, the strongest acousto-optic overlap occurs when phase-matched at the Brillouin scattering condition, which for longitudinal waves in silicon (as is the case in most HBAR-FBAR variants) is $\approx$ 35 GHz. At most other frequencies of interest (e.g., the 3~GHz to 8~GHz range), the acousto-optic interaction strength is dominated by the moving boundary effect, which is very weak, leading to poor optical sideband scattering (i.e., poor upconversion from the acoustic domain to the optical domain) \cite{valle_high-frequency_2019}. One can compensate partially for the low $g_{0}$ by enhancing the parametric interaction strength $G$ through a large intracavity photon number $N_{cav}$, but since $G= g_{0}\sqrt{N_{cav}}$, a linear reduction in $g_{0}$ must be accompanied by a quadratic increase in $N_{cav}$ to realize the same $G$. As we discuss later in Fig.~\ref{fig:g0N}, for existing optomechanical resonators operating at temperatures below 4~K, the values for $G$ achieved in these resonators is on par with that achieved in the 1-D nanobeam optomechanical crystals described in the next section. We note that one can work around the above conditions to a certain extent through use of crystalline substrates with slower acoustic velocity (and lower refractive index) like silica, but the lack of control in engineering strong acousto-optic mode overlap and small mode volume, high-$Q_{o}$ optical cavity geometries is a significant limitation if the ultimate aim is to achieve an overall conversion efficiency that is close to unity at a frequency set by the qubit.

One way to reduce the large optical cavity mode volumes is to engineer bulk wave resonances in an integrated photonic platform. This can be done by engineering a film bulk acoustic wave resonance (FBAR) in a microdisk resonator that modulates an optical whispering gallery mode that propagates around the periphery of the disk \cite{han_cavity_2020}. This approach in principle combines the high $\eta_{PIE}$ of an FBAR-like mode with the high $Q_{o}$ and moderate mode volumes of a microdisk resonator. On the other hand, there are two main challenges that need to be addressed to achieve high transduction efficiencies using this approach. The first is effective actuation of the FBAR breathing mode, given that in contrast to traditional FBAR resonators, here the metals need to be separated from the disk in order to preserve the high $Q_{o}$. Han et al. address this problem by designing a superconducting RF resonator on top of the microdisk to ensure effective excitation of the breathing mode and achieve an electromechanical cooperativity $C_{em}\approx$ 7 \cite{han_cavity_2020}. The second problem is more fundamental in that the $g_{0}$ in these architectures is reduced, especially in comparison to the nanobeam optomechanical crystals discussed below. One way to understand this is to see that the FBAR mode is effectively a thickness mode of the whole microdisk, but only the displacement on the periphery translates to optomechanical coupling, since the optical WGM is localized there. This reduces the effective overlap between the optical and mechanical modes resulting in low $g_{0}$. While we can partially compensate for this with a large $N_{cav}$, the low $g_{0}$ effectively makes it challenging to reach the matching condition ($C_{EM} \approx C_{OM} + 1$) and limits the ultimate achievable transduction efficiency. 

\section{Transduction using high-$g_{0}$ 1D nanobeam optomechanical crystals}

An alternative approach, pursued by many groups worldwide, is to start at the opposite regime. Given that 1D optomechanical crystals (schematic shown in Fig. \ref{fig:Approaches}(b)) have shown strong photon-phonon interactions by confining low-loss mechanical and optical modes in wavelength scale cavities \cite{chan_optimized_2012, ramp_elimination_2018}, provided one can inject phonons efficiently into these cavities from a microwave source, high conversion efficiencies can be achieved. The challenge is that efficient phonon injection here amounts to focusing sound into a nanoscale volume from a microwave transmission line. This poses two competing issues. Mode matching to nanoscale cavities ideally requires transducers (for example, IDTs for launching surface acoustic waves (SAWs)) that are comparable in size to the acoustic cavity. But the electrical impedance of such small transducers is usually in the M$\Omega$ range, making it challenging to impedance match these devices to standard 50 $\Omega$ transmission lines. One can in principle try and work with high impedance circuits, but in general, the efficiency of microwave-to-acoustic wave conversion is limited for nanoscale IDTs (even when impedance matched), on account of the reduced finger overlap \cite{datta1986surface}, which reduces the motional capacitance (alternately, increases the motional impedance significantly). A natural route to avoiding this tradeoff is to work with focusing transducers \cite{siddiqui2018lamb} that can be designed to be impedance matched to 50 $\Omega$ microwave inputs and focus the sound down to nanoscale dimensions. 

\begin{figure*}
\begin{center}
\includegraphics[width=0.92\linewidth]{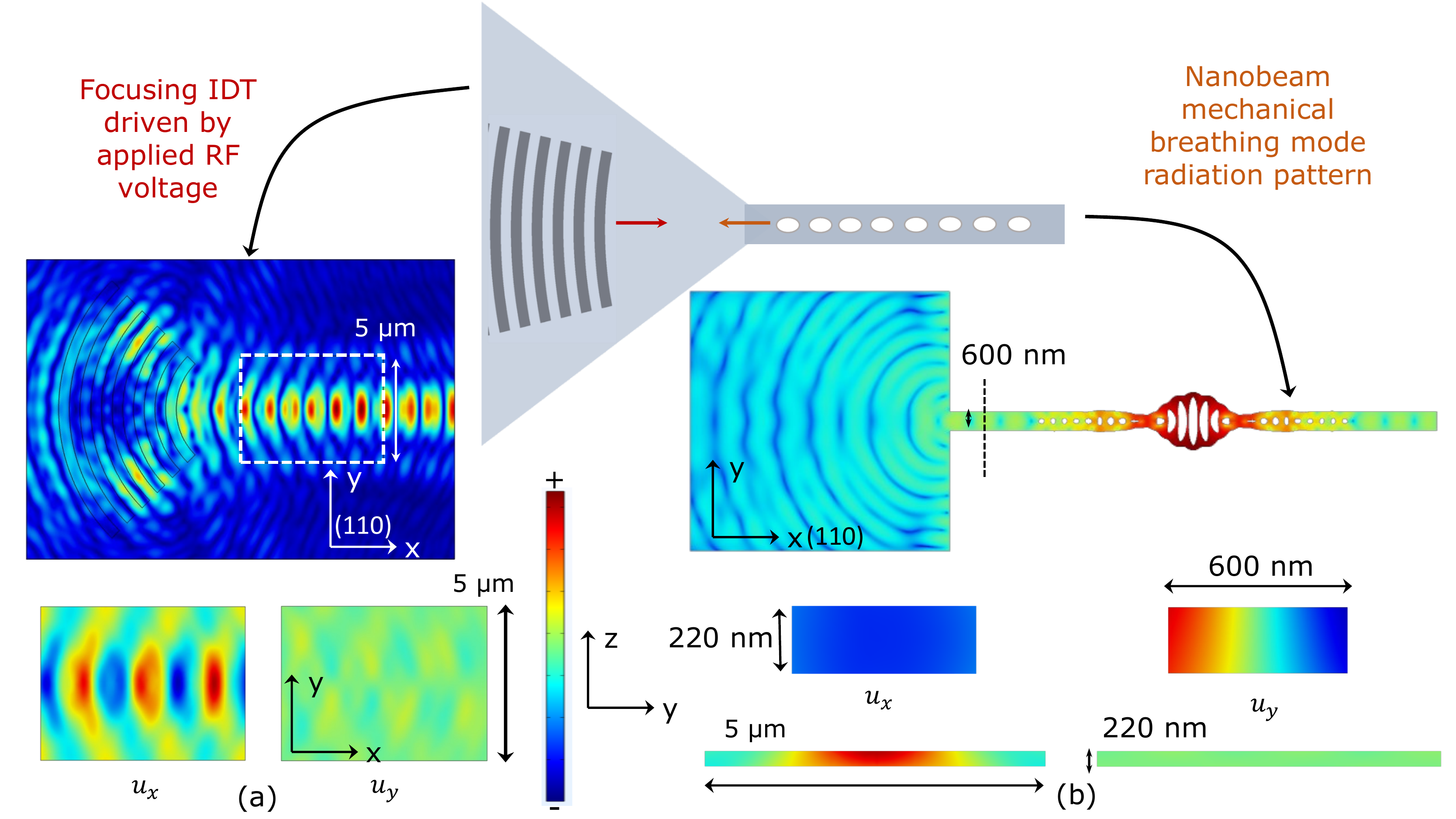}
\end{center}
\caption{Excitation efficiency of nanobeam breathing modes through focusing IDTs. On the left the displacement field (total) of a  focused acoustic beam generated by an Al IDT ($t_{Al} = 50 nm$ on a GaAs membrane ($t_{GaAs} = 220 nm$) is shown. The wave period is chosen to be 2 ${\mu}m$ corresponding to the $S_{0}$ Lamb wave mode in the GaAs membrane at frequency $\approx$ 2.17 GHz. Zoomed-in plots of the x ($u_{x}$) and y ($u_{y}$ displacement fields are shown in the focus region. The color bar range has been kept constant, clearly showing that that $u_{y}$  is significantly lower than $u_{x}$. The right hand side plot shows the displacement (total, log-scale) of an eigenfrequency simulation of a nanobeam breathing mode (f$\approx$ 2.165 GHz). The leakage radiation of the nanobeam mode into the GaAs membrane through cylindrical waves can be clearly seen. The $u_{x}$ and $u_{y}$ displacement components for the breathing mode leaking into the beam are shown on the right. The displacements are plotted at the dashed cut and the colorbar normalized to max($u_{x}$, $u_{y}$). There is significant elastic energy in the $u_{y}$ field, which overlaps poorly with the focused SAW mode. For reference, the $u_{x}$ and $u_{y}$ cross-section components of the acoustic beam generated by the curved IDT are plotted at the focus point (note: the scale bars for the two cross-section plots are different). For efficient excitation (reaching critical coupling), the two displacement fields (nanobeam and focused IDT) should be relatively similar.}
\label{fig:shear wave symmetry}
\end{figure*}

While focusing transducers have been extensively studied \cite{de2003focusing}, there have been very few experimental demonstrations that show that sound can be effectively focused to nanoscale volumes while maintaining high modal purity \cite{siddiqui2018lamb}. A focusing transducer is effectively an acoustic lens that can focus sound down to the (acoustic) diffraction limit. Like the case with designing diffraction-limited optics, the key challenge is overcoming aberrations, in particular acoustic anisotropy \cite{hopcroft2010young} which can be quite significant in most crystalline materials, which are commonly used for their strong opto-mechanical interaction strengths (cf. silicon and GaAs) and high optical and mechanical quality factors. Even if the acoustic anisotropy is compensated, the piezoelectric coefficient is orientation-dependent and goes to zero at certain angles in-plane (ex: along the [100] axis GaAs \cite{de2003focusing}), significantly reducing the conversion efficiency.

Even if efficient focusing transducer geometries can be realized, there is a limit on the excitation efficiency of the optomechanical cavity mode by focusing IDTs that is imposed by the overlap between the two modes.  The highest $g_{0}$ in 1D optomechanical cavities has been traditionally achieved with breathing modes, which are predominantly shear horizontal (SH) modes with an odd $y$-displacement symmetry ($u_{-y}=-u_{+y}$, $u_{y}$ is the displacement field in the $y$ direction) as shown in Fig.~\ref{fig:shear wave symmetry}. The excitation efficiency of the nanobeam mode by a focusing IDT can be quantified by an overlap integral between the IDT displacement field $\vec{u}_{IDT}$ at the focus (assumed to be at the beam entrance) and the displacement field at the beam entrance $\vec{u}_{NB}$ generated by the leaked nanobeam displacement:

\begin{equation}
      \eta = \frac{\int{|\vec{u}^{*}_{IDT}\vec{u}_{NB}|^2\vec{dA}}}{\int{|\vec{u_{IDT}}|^{2}\vec{dA}}\int{|\vec{u_{NB}}|^{2}\vec{dA}}}
      \label{eq:mode_overlap}
\end{equation}

\noindent where the integrals are carried out over the beam cross-section area and $\vec{u}=u_{x}\hat{i}+u_{y}\hat{j}+u_{z}\hat{k}$.

Figure~\ref{fig:shear wave symmetry} plots finite element method simulation results of the acoustic mode generated by a focusing IDT oriented to focus the beam along (110) GaAs and the acoustic radiation generated by a nanobeam breathing mode coupled to a membrane (housing the IDT). The displacement profile generated by the focusing IDT is calculated by driving the IDT at a fixed frequency by applying an oscillating voltage to the electrodes. The nanobeam displacement is calculated from an eigenfrequency simulation, where the nanobeam is attached to a membrane (which would house the focusing IDT, not shown).  The microwave to optical transduction process needs to be bi-directionally efficient, in that the focusing IDT needs to efficiently excite the nanobeam breathing mode (MW to optical), and the IDT also needs to be able to effectively sense the leakage radiation from the breathing mode (optical to MW). To study the overall transduction process, we can focus on the two halves independently. On the left, we show the acoustic beam generated by a focusing IDT. Inspecting the acoustic displacement components in the focus region, we see that the acoustic energy in the $u_{y}$ field is negligible compared to $u_{x}$. On the other hand, the nanobeam breathing mode displacement shows the opposite trend with the acoustic energy primarily residing in the $u_{y}$ field component and the $u_{x}$ component far smaller. As the breathing mode leaks into a beam mode that then radiates into a Lamb wave, by observing the displacement in the beam (dashed line), we can see that the leaked mode still preserves a significant energy in the $u_{y}$ component, which has low overlap with the focusing displacement wave generated by an IDT. This places a limit on the overall (single-pass) transduction efficiency. Given the coupling rate is significantly reduced in this configuration, achieving critical coupling would require very high $Q_{m}$ resonators. While such high-Q devices have been demonstrated \cite{macCabe_phononic_2019}, they come with the inherent trade-off of low transduction bandwidths. Ideally, one would like to increase the resonator to IDT coupling strength in order to operate these devices in an over-coupled regime, which is more suited for high efficiency quantum transduction \cite{wu2020microwave}.     

One way to understand this limitation in overlap efficiency is to recognize that the optomechanical crystals were originally developed and optimized to achieve strong optomechanical coupling strengths with transverse electric (TE) optical mode polarizations. As such the acousto-optic field overlaps are maximized for an electric field that is polarized along $\vec{y}$. From the displacement field of an interdigitated transducer, it is clear that the (RF) electric fields induced by the transducer are predominantly in the x-z plane, which is characteristic of Rayleigh-Lamb waves. The focusing IDT geometry helps overcome the mode mismatch problem from the perspective of size mismatch, but it can not overcome this mode symmetry issue. We want to clarify here that the focused $S_{0}$ Lamb wave mode generated by the IDT has the same overall mode displacement symmetry (in terms of the $u_{x}$, $u_{y}$ and $u_{z}$ field components) as the nanobeam breathing mode, but still has low coupling efficiency because the relative magnitudes of the displacements ($u_{x}$ and $u_{y}$ in particular) are orthogonal, reducing the overlap integral in eq.~\ref{eq:mode_overlap}. One could potentially achieve better overlaps by working with strongly anisotropic materials like lithium niobate, where shear waves can be efficiently excited using IDT-like geometries~\cite{olsson2014high}. However, materials such as lithium niobate have significantly weaker photoelastic coupling coefficients and lower refractive indices than GaAs or Si, resulting in a $\approx$ 10$\times$ reduction in $g_{0}$.

An alternative approach to simultaneously address the impedance and mode mismatch problems that was recently proposed is to use hybridization of localized mechanical modes as a means to link the piezoelectric and optomechanical transducers~\cite{wu2020microwave}. In this scheme, the breathing mode of the 1D nanobeam optomechanical crystal is hybridized with the mechanical mode of a piezoelectric resonator \cite{ju2000detection}. This essentially provides a direct means to piezoelectrically access the nanobeam breathing mode and its co-localized optical mode, without the traveling acoustic wave intermediary used in IDT/SAW-based approaches. However, the mode overlap between the mechanical mode and the optical mode (which determines the strength of the optomechanical interaction) is now reduced relative to an isolated nanobeam optomechanical crystal. This occurs because a part of the mechanical mode is now contained in the piezoelectric resonator which does not overlap with the optical mode located in the nanobeam. Limiting the resulting reduction in $g_{0}$ is important to ensure that large $C_\text{OM}$ is maintained. This can be done using piezoelectric (contour-mode) resonators with a small motional mass, but it comes at the expense of a high electrical impedance. Impedance matching to a 50~$\Omega$ transmission line can be realized through an additional (inductor-capacitor) LC resonator, which has the further benefit of resonantly enhancing the piezoelectric interaction, so that large $C_\text{EM}$ can be realized even with a weak $g_\text{EM}$ \cite{han_cavity_2020}. Reference~\cite{wu2020microwave} predicts that in GaAs, near-unity transduction efficiency and low added noise can be simultaneously achieved using the above approach without requiring any significant advances in $Q_{o}$, $Q_{m}$, $Q_{MW}$, $g_{0}$, and $g_\text{EM}$ beyond what has already been demonstrated in this platform. However, one overriding challenge is that of relying on multiple, coupled resonant elements whose frequencies should align - in this case, the two mechanical resonators and the LC electrical resonator. Such frequency matching can be difficult in practice, in particular, as the mechanical resonators lack the readily available tuning knobs common to integrated photonics. We note that the recent experimental demonstration of superconducting qubit-to-optical photon transduction experiment follows a similar approach in hybridizing the nanobeam breathing mode with a propagating mode in a silicon phononic crystal waveguide, which can be efficiently excited by a piezoelectric AlN layer deposited on top of the silicon ~\cite{mirhosseini2020superconducting}.


\section{Other transducer geometries}

In addition to the two main approaches discussed above, resonant acousto-optic modulators have been demonstrated by placing interdigitated transducers adjacent to microring resonators~\cite{tadesse_acousto-optic_2015,shao2019microwave} and other high $Q_{o}$ cavities, e.g., 2D photonic crystals~\cite{fuhrmann2011dynamic}. These geometries are not ideal for quantum transduction because the conversion efficiencies that can be achieved here are fundamentally limited. The main advantage of working with microring-based geometries is the prospect of achieving high $Q_{o}$ at moderate mode volumes. But the high $Q_{o}$ is not sufficient to compensate for the drawbacks this architecture faces from a signal transduction perspective. The key challenge with any of the microring-based approaches is that the modes that show reasonably high acousto-optic interaction strength, as quantified by $g_{0}$, are the radial breathing modes~\cite{Balram_Moving_2014} which are very challenging to excite with an IDT, and hence the modulation efficiency (and $g_{0}$) that can be achieved is limited. While one can try and engineer a double resonance using the IDT designed as a contour mode resonator (CMR) around an optical micro-ring \cite{shao2019microwave}, most of the acoustic energy lies within the contour mode (under the IDTs) and the effective displacement participation ratio is small. One can look at this inefficiency in two equivalent ways. First, the overall ${\eta}_{PIE}$ is small because most of the injected phonons are not contributing to modulation. Alternately, $g_{0}$ is reduced because the effective mass of the mechanical mode has been significantly increased by the CMR. We note that, apart from radial breathing modes, microdisk film thickness modes can be piezoelectrically excited in an FBAR geometry, as noted in Section III and demonstrated in Ref.~\cite{han_cavity_2020}, at the expense of a weaker $g_{0}$. 

While most of the effects discussed in this paper are fundamental in nature, there are some practical issues to consider as well. Fabrication processes for most of the nanoscale quantum transducers require multiple layers of (aligned) lithography and patterning and overlay errors between the different layers will result in large device-to-device variation. For example, consider a microring resonator that is excited by an IDT. The actual acoustic energy inside the microring is very sensitive to both the relative position of the IDT with respect to the ring and the magnitude of the undercut, assuming the microring is suspended~\cite{jiang2020suspended}. Such overlay errors will also affect the other geometries that are being considered in this work, but it is magnified for the rings because we are not working with a discrete mechanical resonance, but instead with broadband (limited by the IDT bandwidth) mechanical excitation of the ring. It is important to keep in mind that state of the art CMOS foundries can achieve overlay errors less than 2 nm on 300 mm wafers, so the overlay errors in research can be attributed primarily to working with small sample sizes ($<$ 1 cm) in non-silicon platforms like GaAs and lithium niobate on insulator.  

\section{Current transducer performance}
\begin{figure}
\begin{center}
\includegraphics[width=1\linewidth]{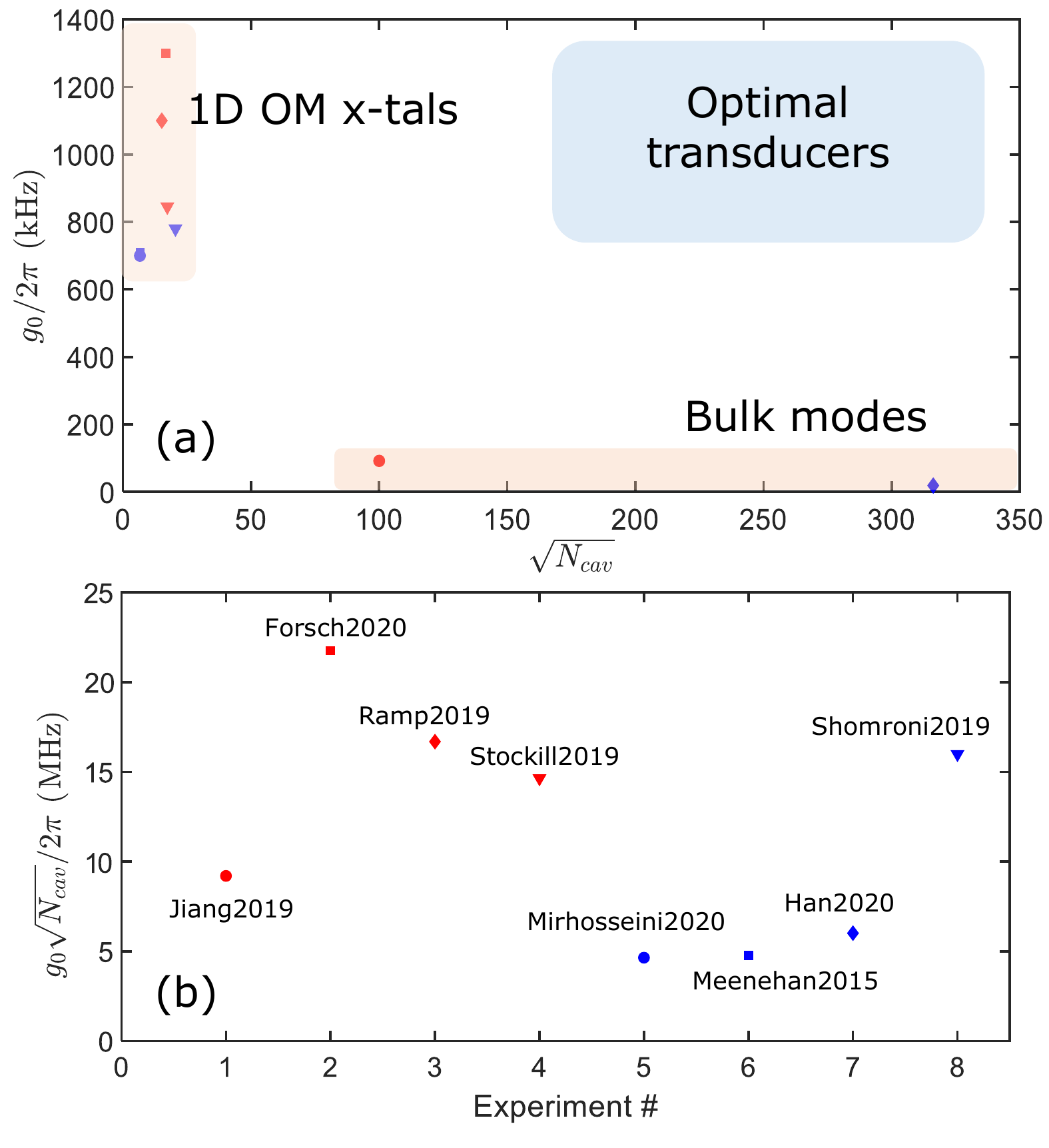}
\end{center}
\caption{(a) The optomechanical coupling strength ($g_{0}$) and intracavity photon number ($N_{cav}$) for a variety of cryogenic ($<$ 4K) optomechanical experiments are plotted. The experiments are clustered in either the high coupling strength $g_{0}$, low intracavity photon number $N_{cav}$ regime for 1D optomechanical crystals (1D OM x-tals) or the low $g_{0}$, high $N_{cav}$ space (bulk modes). Optimal transduction requires simultaneously high $g_{0}$ and $N_{cav}$. (b) $g_{0}\sqrt{N_{cav}}$, the square of which is proportional to the optomechanical cooperativity $C_\text{OM}$, is within a factor of 3 for all current platforms.}
\label{fig:g0N}
\end{figure}

\begin{table*}[]
\begin{tabular}{|c|c|c|c|c|c|c|c|c|}
    
 \hline
 Material & Temperature & $g_{0}/2\pi$ (kHz) & $N_{cav}$ & $C_{om} = 4g_{0}^{2}/\kappa\gamma$ & $\eta_{PIE}$ & $\eta_{peak}$  & $n_{th}$ & Ref. \\
 \hline
 LiNbO$_3$  & $\approx$ 4 K  & 92 & $10^4$ & $1.5*10^{-4}$ & $10^{-3}$ & $10^{-5}$ & 45.29 & Jiang2020* \cite{jiang_lithium_2019}\\
 \hline
 GaAs & 20 mK to 100 mK & 1300 & 280 & $5.9*10^{-3}$ & $3.6*10^{-10}$ & $5.5*10^{-12}$ & 0.755 & Forsch2020 \cite{forsch2020microwave} \\
\hline
AlN/Si hybrid & 20 mK to 100 mK & 420 & 44 & $-$ & - & $1.5*10^{-5}$ & 0.41 & Mirhosseini2020 \cite{mirhosseini2020superconducting} \\
\hline
AlN & 20 mK to 100 mK & 19 & $10^5$ & $4*10^{-6}$ & $\approx1$ & $7.3*10^{-4}$ & 0.21 & Han2020 \cite{han_cavity_2020} \\
\hline
\end{tabular}

\caption{List of current (cryogenic) microwave-to-optical transduction experiments in a variety of material platforms and geometries, with each row summarizing the various relevant experimental parameters. We note that experiments at 4K benefit from a significantly larger cryogenic cooling capacity compared to experiments in a dilution fridge, impacting $N_{cav}$. *The LiNbO$_3$ experimental parameters are taken from both \cite{jiang_lithium_2019, jiang_efficient_2020}. Reference \cite{jiang_lithium_2019} carried out the transduction at 4K, while Ref.~\cite{jiang_efficient_2020} improves the design and should see improvement in cryogenic transducer performance as well. The AlN experiment in Ref.~\cite{han_cavity_2020} uses a microwave cavity to realize $C_\text{EM}\approx7$, the largest electromechanical cooperativity thus far achieved in an MW-OT. }
\label{table: MW to Optical current parameters}
\end{table*}

As discussed in the previous sections, both the high $g_{0}$, low $N_{cav}$, low $\eta_{PIE}$ 1D nanobeam optomechanical crystal approach and the low $g_{0}$, high $N_{cav}$, high $\eta_{PIE}$ bulk acoustic wave resonator approach towards quantum microwave-to-optical signal transduction have (complementary) limitations. One way to understand these limitations is to look at Fig.~\ref{fig:g0N}(a), which plots the single photon optomechanical coupling strength $g_{0}/2\pi$ as a function of intracavity pump photon number $\sqrt{N_{cav}}$ for selected cryogenic ($<$ 4K) optomechanical experiments. All the experiments fall into one of two broad classes: the 1D optomechanical crystal (1D OM x-tals) variants show high $g_{0}/2\pi$, but $N_{cav}$ is low due to high surface absorption, which either causes a thermo-optic instability or for ultra-low temperature experiments ($\approx 100$~mK), can heat the system out of the mechanical ground state. On the other hand, bigger cavity geometries (labeled as bulk modes in Fig.~\ref{fig:g0N})(a)) support large $N_{cav}$ at the cost of lower $g_{0}/2\pi$. Since the optomechanical cooperativity scales quadratically with $G=g_{0}\sqrt{N}$, both approaches achieve figures of merit that are within a factor of 3 of each other as shown in Fig.~\ref{fig:g0N}(b); e.g., Ref.~\cite{forsch2020microwave} for nanobeam optomechanical crystals and Ref.~\cite{han_cavity_2020} for bulk modes. To achieve optimal transduction, ideally we need devices that can work in the top right corner of Fig.~\ref{fig:g0N}(a), supporting both large $g_{0}$ and large $N_{cav}$ simultaneously. On the 1D optomechanical crystal front, higher $N_{cav}$ can be achieved by improving surface passivation techniques \cite{najer2019gated} to reduce the significant surface absorption effects inherent in large surface to volume ratio photonic crystal cavities. Improving $g_{0}$ in bulk cavities is much harder as it mainly relies on engineering cavities in materials like tellurium dioxide which have stronger acousto-optic interactions than typical platforms used for piezoelectric optomehanics, by virtue of having larger photoelastic coefficients. A related problem with bulk resonators is that highest $g_{0}$ values are usually achieved at the Brillouin frequency, which fixes the frequency of operation and does not allow the mechanical frequency tunability inherent in nanoscale geometries. 

Table \ref{table: MW to Optical current parameters} summarizes the relevant experimental parameters for state-of-the art cryogenic transduction experiments carried out in a variety of materials and device platforms. While the experiments in Refs.~\cite{jiang_efficient_2020, mirhosseini2020superconducting, forsch2020microwave} use an IDT to excite the breathing mode of a 1D optomechanical crystal, the experiment in Ref.~\cite{han_cavity_2020} uses a high-frequency FBAR mode of AlN, coupled to a microwave cavity, to modulate the whispering gallery optical mode supported by the AlN microdisk. As can be seen from the $\eta_{peak}$ values in the table, state-of-the art experiments using both the nanobeam and HBAR/FBAR routes achieve $\approx10^{-5}$ in photon transduction efficiency, although their limitations are complementary as discussed. The nanobeam experiments are mostly limited on the electromechanical front ($\eta_{PIE}$) whereas the (quasi) bulk resonators are mostly limited by the achievable $g_{o}$ and $C_\text{OM}$. Given these limitations, one is naturally inclined to ask whether there exist novel geometries that sit somewhere between these two extremes and can achieve significantly higher transduction efficiency than either one (i.e., the optimal transducers represented by the shaded blue region in Fig.~\ref{fig:g0N}(a)). The co-design approach, referred to in the title and discussed in more detail later in Section VIII, is one systematic way to design better quantum transducers.

\section{Material platforms: monolithic vs hybrid}

Before moving on to the co-design approach, we would like to discuss the choice of material platform for implementing quantum transducers. This problem has been considered before~\cite{wu2020microwave} and while there is no one material platform that is clearly optimal, there are some general requirements that are helpful to heed. In particular, optimal transduction requires matching of the electromechanical and optomechancial cooperativities and most monolithic approaches favor one over the other. For example, materials like GaAs and Si provide an easier path to large $C_\text{OM}$ due to their large photoelastic coefficients and refractive indices, but have weak (GaAs) or no (Si) piezoelectric coupling. Materials like LiNbO$_3$ and AlN have large piezoelectric coupling but weaker photoelastic coefficients and smaller refractive indices, making large $C_\text{OM}$ hard to realize without large $N_{cav}$. From these considerations, one might naturally conclude that hybrid material platforms, e.g., AlN-on-Si or LiNbO$_3$-on-Si, are the most promising path for simultaneously realizing large $C_\text{EM}$ and large $C_\text{OM}$. 

While such a hybrid approach is promising and is indeed being pursued by multiple groups, for example, using AlN-on-Si in the recent demonstraton of qubit-to-optical photon transduction~\cite{mirhosseini2020superconducting}, it is important to keep in mind that engineering high efficiencies in hybrid phononic platforms is much more challenging than in similar hybrid photonic platforms. This comes about mainly because while the photonic platform can be designed to be effectively single mode and one can engineer high-efficiency transitions between the different layers through simple evanescent coupling via adiabatic tapers, the phononic platform is inherently multi-mode and every transition between the two materials provides a path for intermodal scattering and loss and a substantial decrease in transduction efficiency. That being said, there are active efforts in combining materials like LiNbO$_3$ and Si \cite{van2020piezo,marinkovic2021hybrid}, where if such acoustic matching issues are properly addressed, significant advances in transduction may be realized. We note that the coupled resonator approach described at the end of Section IV, in which the mechanical modes of a piezoelectric resonator and an optomechanical resonator are hybridized to enable efficient microwave and optical access, may provide one compelling approach to address these acoustic mode and impedance matching challenges.

\section{Co-design: Transverse magnetic mode optomechanical crystals}

\begin{figure*}
\begin{center}
\includegraphics[width=0.8\linewidth]{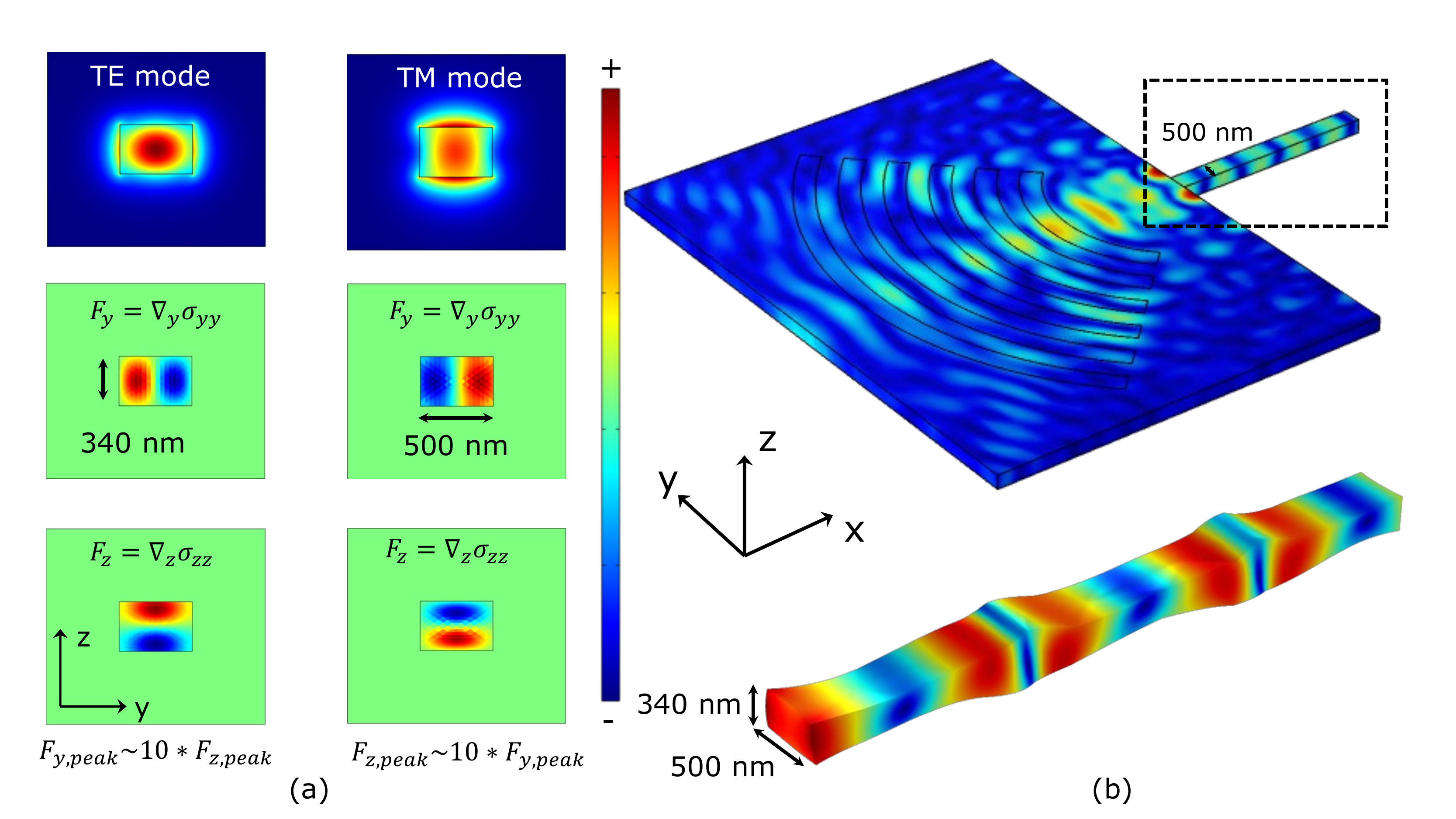}
\end{center}
\caption{Contrasting the dominant electrostrictive optical forces induced by propagating transverse electric (TE) and transverse magnetic (TM) fields in a silicon waveguide. The TE mode predominantly induces x-directed forces, whereas the TM mode induced y-directed forces. The right hand side shows a focusing acoustic wave, excited by an IDT, in a GaAs membrane (t = 340 nm) incident on a beam (width = 500 nm, t = 340 nm). The zoomed-in displacement of the beam is shown below for reference. Given the displacement profile induced by the focusing IDT at the beam entrance, a shear vertical-like (SV) beam mode is expected to have a higher excitation efficiency than a shear horizontal (SH) mode. }
\label{fig:co-design}
\end{figure*}

The co-design approach, advocated in the title, refers to a methodology where an overall figure-of-merit is maximized, subject to a simultaneous multi-domain optimization. To be concrete, an optimal quantum transducer would be designed to maximize $\eta_{peak}$ (eq.~\ref{eq:peakEta}). Given that multiphysics softwares like COMSOL \footnote{Certain commercial products or company names are identified here to describe our study adequately. Such identification is not intended to imply recommendation or endorsement by the National Institute of Standards and Technology, nor is it intended to imply that the products or names identified are necessarily the best available for the purpose.} can simultaneously solve the microwave, mechanics and optics parts of the problem, it opens up the design space for considering alternative geometries that are specifically tailored for the transducer application. In particular, as discussed above, existing approaches to the problem have limitations that originate from the fact that these structures (both the 1D optomechanical crystals and HBAR resonators) were originally designed for a different purpose (cavity optomechanics and RF filters respectively), without specific consideration of the challenges of quantum transduction, and as we have outlined above, that imposes significant challenges in realizing high $\eta_{peak}$.

As an example of the co-design approach, we revisit the overlap integral between the IDT displacement field and the nanobeam breathing mode in one-dimensional optomechanical crystals. One way to think of this overlap problem is that we want to maximize the overlap between the electrostrictive forces exerted by the optical field in the cavity and the displacement forces exerted by the focusing IDT. Provided the same mechanical mode is driven in the two cases, you can guarantee high excitation efficiency.

Figure~\ref{fig:co-design} shows the electrostrictive forces exerted by a transverse electric (TE) polarized optical mode propagating in the waveguide (width=500 nm, thickness=340 nm). We chose to work with a waveguide mode instead of the photonic crystal cavity mode for the main reason that the symmetry arguments are easier to visualize. As expected from our previous discussion, the TE mode predominantly exerts \(y\)-forces with the \(z\)-directed forces being an order of magnitude smaller. Figure~\ref{fig:co-design} (b) shows the (total) displacement field for a focused Lamb wave mode generated by an IDT incident on the same beam. The zoomed-in displacement in the beam is plotted below for reference. It is clear that the beam mode retains the shear vertical (SV-like) character of the incident $S_{0}$ Lamb wave mode, which is predominantly displaced in the \(x-z\) plane, characteristic of Rayleigh-Lamb like mode. We note that within the beam, it is not strictly correct to refer to the modes as shear horizontal or vertical as those definitions strictly apply when there is no confinement along \(y\). The overlap integral, described in eq. \ref{eq:mode_overlap} can be reframed as related to the efficiently with which this beam mode be excited by a TE polarized waveguide mode. Given the smaller overlap between the strong \(x\)-directed force and lower \(x\)-directed displacement, this transfer of energy is poor. Alternately, the beam mode should have a shear-horizontal (SH-like) displacement for efficient excitation by a TE mode.

On the other hand, if we instead consider a TM polarized waveguide mode, the forces are now primarily \(y\)-directed and in phase with the large \(y\)-displacement of the waveguide mode. This means that the TM mode is more suited to driving the (SV-like) Lamb wave modes of the beam. Traditionally, optomechanical crystals, and photonic crystals in general, have been designed to work with TE polarized optical fields. At least in silicon, this requirement is mainly due to 220 nm being the dominant silicon device layer thickness, and it supports only a TE mode. On the other hand, increasing the silicon thickness to 340 nm, a low loss TM mode can now be supported and high $Q_{o}$ optomechanical crystal cavities can be designed around it. While these structures might not have the same $Q_{o}$ as the TE modes, the $g_0$ should be similar, given the mode index is very close and the cavity mode volumes are nominally similar. More importantly, they will be well-coupled to the beam modes, which can then be driven and read out by a focusing IDT structure, with the prospect of achieving critical coupling, without resorting to ultra-high $Q_{m}$. While significant design work, not to mention experiments, are needed to demonstrate the benefits of this geometry in practice, the basic physical arguments presented here indicate the intrinsic value of co-design when developing piezo-optoemchanical transducers.

\section{Off-chip filtering}

\begin{figure}
\begin{center}
\includegraphics[width=1\linewidth]{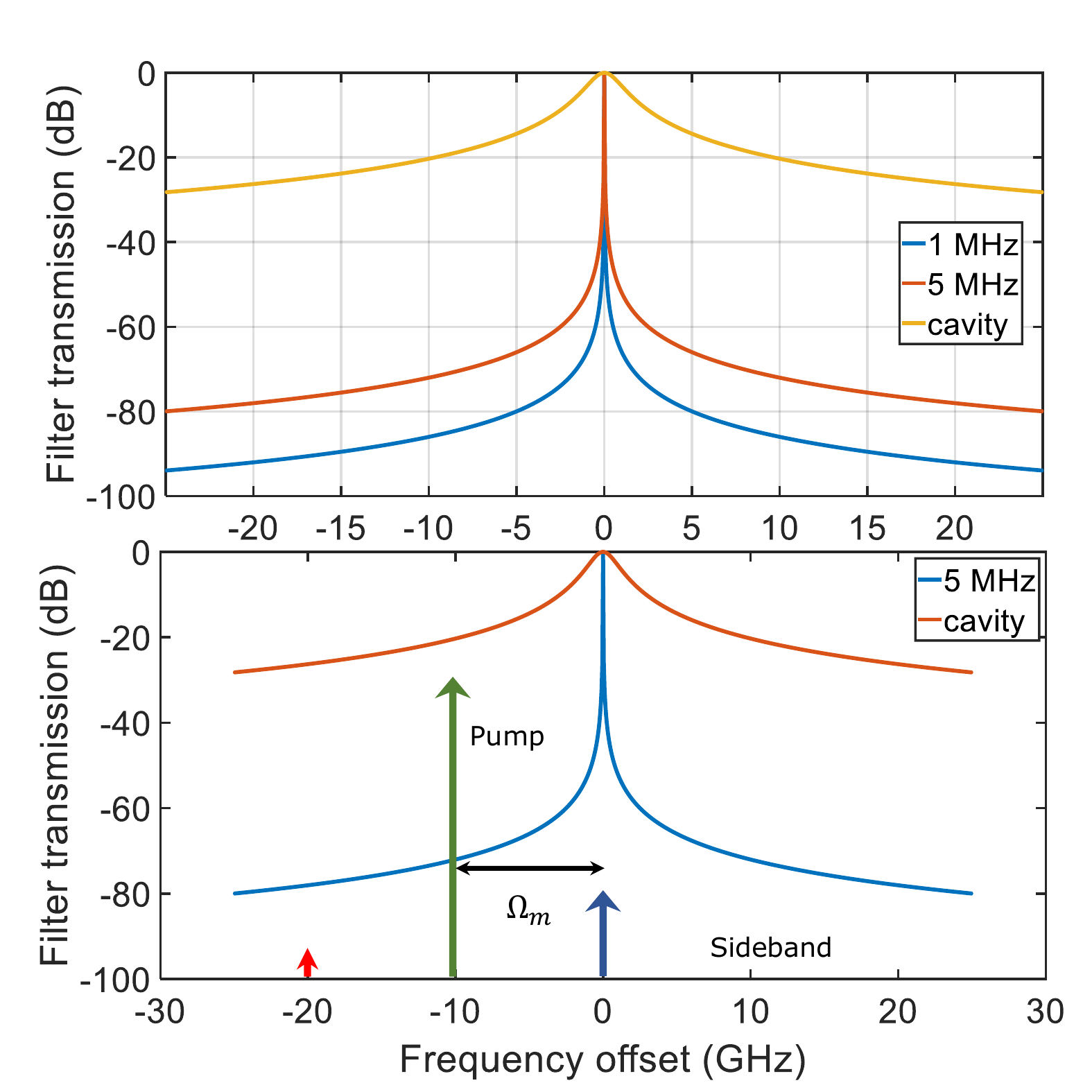}
\end{center}
\caption{Pump suppression that can be achieved using external fiber Fabry-Perot filters with bandwidths of 1~MHz (blue) and 5~MHz (red) added after the optomechanical cavity whose optical linewidth is 1.92~GHz ($Q_{o}=10^5$ at 1550~nm). Working with high mechanical mode frequencies (10 GHz) allows effective pump filtering. Moving from 2.5 GHz mechanical modes to 10 GHz mechanical modes provides an additional 10 dB of pump suppression. }
\label{fig:filtering}
\end{figure}


As noted in Section II, the mechanical frequency plays a particularly important role in the context of added noise in the quantum transducer.  Higher mechanical frequencies are generally preferred from the standpoint of both a lower thermal occupancy at a given temperature, and a greater amount of sideband resolution for a given optical cavity linewidth. Of course, in practice, this mechanical frequency is likely to be chosen not purely based on noise considerations, but instead will be linked to the superconducting qubit frequency, unless an intermediate stage of microwave-to-microwave quantum frequency conversion is employed~\cite{lecocq_mechanically_2016}. Given that this likely means that the amount of sideband resolution is finite, one of the key requirements for enabling quantum state transduction is efficient pump-suppression. The optical pump is used to parametrically enhance the optomechanical interaction, but couples out from the cavity in the same spatial mode as the sideband. In this case, the pump effectively serves as source of noise (e.g., in characterizing the photon statistics of the sideband) and needs to be filtered sufficiently. The cavity provides some degree of pump suppression, but most quantum protocols require high fidelities which can only be achieved with external fiber Fabry-Perot filters. Figure~\ref{fig:filtering} provides an illustration of the pump suppression provided by the optomechanical cavity ($Q_{o}\approx 10^5$) and the external Fabry-Perot filter.

The requirement of sufficiently filtering the pump also provides an interesting perspective on the choice of the transducer. As discussed above, the two complementary approaches on using high $g_{0}$, low $N_{cav}$ 1D nanobeam-like cavities and low $g_{0}$, high $N_{cav}$ bulk HBAR like devices present complementary challenges as can be seen from a simple calculation. An estimate of the degree of pump suppression required can be obtained from analyzing the transducer output channel for sideband to pump photon signal-to-noise ratio of 1. This means that for every sideband photon present, we can tolerate at maximum one photon from the pump channel spectrally leaking into our detector. To estimate this, we note that effective transduction requires an optomechanical cooperativity $C_{om}\approx$ 1. To achieve this in a 1D optomechanical crystal with $g_{0}/2\pi$=1~MHz, $Q_{m}=10^4$, and $Q_{o}=10^5$ requires a pump suppression of -25 dB. On the other hand, with an HBAR-like cavity with nominal parameters  $g_{0}/2\pi=1$~kHz, $Q_{m}=10^5$, and $Q_{o}=10^6$ requires a pump suppression of -65 dB. The stark difference points to the importance of high $g_{0}$. Working with an HBAR-like configuration makes it imperative to work with very high frequency mechanical modes ($\approx$ 10 GHz) to achieve the necessary pump filtering with an external Fabry-Perot cavity. While the 1D nanobeam transducers can work with lower frequency mechanical modes (2.5~GHz to 5~GHz), moving to higher mechanical frequencies while maintaining $g_{0}$ is critical to meeting more stringent SNR requirements.

\section{Conclusions}

In this perspective, we have discussed some of the main challenges and opportunities facing piezoelectric routes towards efficient quantum microwave-to-optical signal transduction. Our main emphasis has been on exploring the underlying device architectures, both the optomechanics-inspired 1D nanobeam cavities and the RF MEMS inspired bulk wave resonators, and analyzing their complementary strengths and weaknesses. Effective transduction requires matched electromechanical and optomechanical cooperativities.  The nanobeam optomechanical cavities and RF MEMS bulk wave resonators operate in different regions of the overall transducer parameter space, but to date have realized similar overall efficiency levels. Ultimately, the need for cooperativity matching points to the importance of co-designing both the microwave-acoustic and acousto-optic subsystems in tandem with the overall transduction efficiency as a key metric. To that end, we have suggested the use of TM-polarized optical modes in 1D nanobeam cavities as a starting point geometry for a more tailored approach to piezo-optomechanical quantum transduction.

\section{Acknowledgements}

KCB would like to thank the European Research Council for funding support (ERC-StG, SBS3-5 758843). KS thanks the ARO/LPS CQTS program for funding support.

\bibliography{NIST-nanophotonics}

\begin{thebibliography}{62}%
\makeatletter
\providecommand \@ifxundefined [1]{%
 \@ifx{#1\undefined}
}%
\providecommand \@ifnum [1]{%
 \ifnum #1\expandafter \@firstoftwo
 \else \expandafter \@secondoftwo
 \fi
}%
\providecommand \@ifx [1]{%
 \ifx #1\expandafter \@firstoftwo
 \else \expandafter \@secondoftwo
 \fi
}%
\providecommand \natexlab [1]{#1}%
\providecommand \enquote  [1]{``#1''}%
\providecommand \bibnamefont  [1]{#1}%
\providecommand \bibfnamefont [1]{#1}%
\providecommand \citenamefont [1]{#1}%
\providecommand \href@noop [0]{\@secondoftwo}%
\providecommand \href [0]{\begingroup \@sanitize@url \@href}%
\providecommand \@href[1]{\@@startlink{#1}\@@href}%
\providecommand \@@href[1]{\endgroup#1\@@endlink}%
\providecommand \@sanitize@url [0]{\catcode `\\12\catcode `\$12\catcode
  `\&12\catcode `\#12\catcode `\^12\catcode `\_12\catcode `\%12\relax}%
\providecommand \@@startlink[1]{}%
\providecommand \@@endlink[0]{}%
\providecommand \url  [0]{\begingroup\@sanitize@url \@url }%
\providecommand \@url [1]{\endgroup\@href {#1}{\urlprefix }}%
\providecommand \urlprefix  [0]{URL }%
\providecommand \Eprint [0]{\href }%
\providecommand \doibase [0]{https://doi.org/}%
\providecommand \selectlanguage [0]{\@gobble}%
\providecommand \bibinfo  [0]{\@secondoftwo}%
\providecommand \bibfield  [0]{\@secondoftwo}%
\providecommand \translation [1]{[#1]}%
\providecommand \BibitemOpen [0]{}%
\providecommand \bibitemStop [0]{}%
\providecommand \bibitemNoStop [0]{.\EOS\space}%
\providecommand \EOS [0]{\spacefactor3000\relax}%
\providecommand \BibitemShut  [1]{\csname bibitem#1\endcsname}%
\let\auto@bib@innerbib\@empty
\bibitem [{\citenamefont {Monroe}\ and\ \citenamefont
  {Kim}(2013)}]{monroe2013scaling}%
  \BibitemOpen
  \bibfield  {author} {\bibinfo {author} {\bibfnamefont {C.}~\bibnamefont
  {Monroe}}\ and\ \bibinfo {author} {\bibfnamefont {J.}~\bibnamefont {Kim}},\
  }\bibfield  {title} {\bibinfo {title} {Scaling the ion trap quantum
  processor},\ }\href@noop {} {\bibfield  {journal} {\bibinfo  {journal}
  {Science}\ }\textbf {\bibinfo {volume} {339}},\ \bibinfo {pages} {1164}
  (\bibinfo {year} {2013})}\BibitemShut {NoStop}%
\bibitem [{\citenamefont {Bardin}\ \emph {et~al.}(2021)\citenamefont {Bardin},
  \citenamefont {Slichter},\ and\ \citenamefont
  {Reilly}}]{bardin2021microwaves}%
  \BibitemOpen
  \bibfield  {author} {\bibinfo {author} {\bibfnamefont {J.~C.}\ \bibnamefont
  {Bardin}}, \bibinfo {author} {\bibfnamefont {D.~H.}\ \bibnamefont
  {Slichter}},\ and\ \bibinfo {author} {\bibfnamefont {D.~J.}\ \bibnamefont
  {Reilly}},\ }\bibfield  {title} {\bibinfo {title} {Microwaves in quantum
  computing},\ }\href@noop {} {\bibfield  {journal} {\bibinfo  {journal} {IEEE
  Journal of Microwaves}\ }\textbf {\bibinfo {volume} {1}},\ \bibinfo {pages}
  {403} (\bibinfo {year} {2021})}\BibitemShut {NoStop}%
\bibitem [{\citenamefont {Wang}\ \emph {et~al.}(2018)\citenamefont {Wang},
  \citenamefont {Paesani}, \citenamefont {Ding}, \citenamefont {Santagati},
  \citenamefont {Skrzypczyk}, \citenamefont {Salavrakos}, \citenamefont {Tura},
  \citenamefont {Augusiak}, \citenamefont {Mančinska}, \citenamefont {Bacco},
  \citenamefont {Bonneau}, \citenamefont {Silverstone}, \citenamefont {Gong},
  \citenamefont {Acín}, \citenamefont {Rottwitt}, \citenamefont {Oxenløwe},
  \citenamefont {O’Brien}, \citenamefont {Laing},\ and\ \citenamefont
  {Thompson}}]{wang_multidimensional_2018}%
  \BibitemOpen
  \bibfield  {author} {\bibinfo {author} {\bibfnamefont {J.}~\bibnamefont
  {Wang}}, \bibinfo {author} {\bibfnamefont {S.}~\bibnamefont {Paesani}},
  \bibinfo {author} {\bibfnamefont {Y.}~\bibnamefont {Ding}}, \bibinfo {author}
  {\bibfnamefont {R.}~\bibnamefont {Santagati}}, \bibinfo {author}
  {\bibfnamefont {P.}~\bibnamefont {Skrzypczyk}}, \bibinfo {author}
  {\bibfnamefont {A.}~\bibnamefont {Salavrakos}}, \bibinfo {author}
  {\bibfnamefont {J.}~\bibnamefont {Tura}}, \bibinfo {author} {\bibfnamefont
  {R.}~\bibnamefont {Augusiak}}, \bibinfo {author} {\bibfnamefont
  {L.}~\bibnamefont {Mančinska}}, \bibinfo {author} {\bibfnamefont
  {D.}~\bibnamefont {Bacco}}, \bibinfo {author} {\bibfnamefont
  {D.}~\bibnamefont {Bonneau}}, \bibinfo {author} {\bibfnamefont {J.~W.}\
  \bibnamefont {Silverstone}}, \bibinfo {author} {\bibfnamefont
  {Q.}~\bibnamefont {Gong}}, \bibinfo {author} {\bibfnamefont {A.}~\bibnamefont
  {Acín}}, \bibinfo {author} {\bibfnamefont {K.}~\bibnamefont {Rottwitt}},
  \bibinfo {author} {\bibfnamefont {L.~K.}\ \bibnamefont {Oxenløwe}}, \bibinfo
  {author} {\bibfnamefont {J.~L.}\ \bibnamefont {O’Brien}}, \bibinfo {author}
  {\bibfnamefont {A.}~\bibnamefont {Laing}},\ and\ \bibinfo {author}
  {\bibfnamefont {M.~G.}\ \bibnamefont {Thompson}},\ }\bibfield  {title}
  {\bibinfo {title} {Multidimensional quantum entanglement with large-scale
  integrated optics},\ }\href@noop {} {\bibfield  {journal} {\bibinfo
  {journal} {Science}\ }\textbf {\bibinfo {volume} {360}},\ \bibinfo {pages}
  {285} (\bibinfo {year} {2018})}\BibitemShut {NoStop}%
\bibitem [{\citenamefont {Kurizki}\ \emph {et~al.}(2015)\citenamefont
  {Kurizki}, \citenamefont {Bertet}, \citenamefont {Kubo}, \citenamefont
  {M{\o}lmer}, \citenamefont {Petrosyan}, \citenamefont {Rabl},\ and\
  \citenamefont {Schmiedmayer}}]{kurizki2015quantum}%
  \BibitemOpen
  \bibfield  {author} {\bibinfo {author} {\bibfnamefont {G.}~\bibnamefont
  {Kurizki}}, \bibinfo {author} {\bibfnamefont {P.}~\bibnamefont {Bertet}},
  \bibinfo {author} {\bibfnamefont {Y.}~\bibnamefont {Kubo}}, \bibinfo {author}
  {\bibfnamefont {K.}~\bibnamefont {M{\o}lmer}}, \bibinfo {author}
  {\bibfnamefont {D.}~\bibnamefont {Petrosyan}}, \bibinfo {author}
  {\bibfnamefont {P.}~\bibnamefont {Rabl}},\ and\ \bibinfo {author}
  {\bibfnamefont {J.}~\bibnamefont {Schmiedmayer}},\ }\bibfield  {title}
  {\bibinfo {title} {Quantum technologies with hybrid systems},\ }\href@noop {}
  {\bibfield  {journal} {\bibinfo  {journal} {Proceedings of the National
  Academy of Sciences}\ }\textbf {\bibinfo {volume} {112}},\ \bibinfo {pages}
  {3866} (\bibinfo {year} {2015})}\BibitemShut {NoStop}%
\bibitem [{\citenamefont {Regal}\ and\ \citenamefont
  {Lehnert}(2011)}]{regal_cavity_2011}%
  \BibitemOpen
  \bibfield  {author} {\bibinfo {author} {\bibfnamefont {C.~A.}\ \bibnamefont
  {Regal}}\ and\ \bibinfo {author} {\bibfnamefont {K.~W.}\ \bibnamefont
  {Lehnert}},\ }\bibfield  {title} {\bibinfo {title} {From cavity
  electromechanics to cavity optomechanics},\ }\href@noop {} {\bibfield
  {journal} {\bibinfo  {journal} {J. Phys. Conf. Ser.}\ }\textbf {\bibinfo
  {volume} {264}},\ \bibinfo {pages} {012025} (\bibinfo {year}
  {2011})}\BibitemShut {NoStop}%
\bibitem [{\citenamefont {Safavi-Naeini}\ and\ \citenamefont
  {Painter}(2011)}]{safavi-naeini_proposal_2011}%
  \BibitemOpen
  \bibfield  {author} {\bibinfo {author} {\bibfnamefont {A.~H.}\ \bibnamefont
  {Safavi-Naeini}}\ and\ \bibinfo {author} {\bibfnamefont {O.}~\bibnamefont
  {Painter}},\ }\bibfield  {title} {\bibinfo {title} {Proposal for an
  optomechanical traveling wave phonon–photon translator},\ }\href
  {https://doi.org/10.1088/1367-2630/13/1/013017} {\bibfield  {journal}
  {\bibinfo  {journal} {New Journal of Physics}\ }\textbf {\bibinfo {volume}
  {13}},\ \bibinfo {pages} {013017} (\bibinfo {year} {2011})}\BibitemShut
  {NoStop}%
\bibitem [{\citenamefont {Lauk}\ \emph {et~al.}(2020)\citenamefont {Lauk},
  \citenamefont {Sinclair}, \citenamefont {Barzanjeh}, \citenamefont {Covey},
  \citenamefont {Saffman}, \citenamefont {Spiropulu},\ and\ \citenamefont
  {Simon}}]{lauk2020perspectives}%
  \BibitemOpen
  \bibfield  {author} {\bibinfo {author} {\bibfnamefont {N.}~\bibnamefont
  {Lauk}}, \bibinfo {author} {\bibfnamefont {N.}~\bibnamefont {Sinclair}},
  \bibinfo {author} {\bibfnamefont {S.}~\bibnamefont {Barzanjeh}}, \bibinfo
  {author} {\bibfnamefont {J.~P.}\ \bibnamefont {Covey}}, \bibinfo {author}
  {\bibfnamefont {M.}~\bibnamefont {Saffman}}, \bibinfo {author} {\bibfnamefont
  {M.}~\bibnamefont {Spiropulu}},\ and\ \bibinfo {author} {\bibfnamefont
  {C.}~\bibnamefont {Simon}},\ }\bibfield  {title} {\bibinfo {title}
  {Perspectives on quantum transduction},\ }\href@noop {} {\bibfield  {journal}
  {\bibinfo  {journal} {Quantum Science and Technology}\ }\textbf {\bibinfo
  {volume} {5}},\ \bibinfo {pages} {020501} (\bibinfo {year}
  {2020})}\BibitemShut {NoStop}%
\bibitem [{\citenamefont {Chu}\ and\ \citenamefont
  {Groeblacher}(2020)}]{chu2020perspective}%
  \BibitemOpen
  \bibfield  {author} {\bibinfo {author} {\bibfnamefont {Y.}~\bibnamefont
  {Chu}}\ and\ \bibinfo {author} {\bibfnamefont {S.}~\bibnamefont
  {Groeblacher}},\ }\bibfield  {title} {\bibinfo {title} {A perspective on
  hybrid quantum opto-and electromechanical systems},\ }\href@noop {}
  {\bibfield  {journal} {\bibinfo  {journal} {Applied Physics Letters}\
  }\textbf {\bibinfo {volume} {117}},\ \bibinfo {pages} {150503} (\bibinfo
  {year} {2020})}\BibitemShut {NoStop}%
\bibitem [{\citenamefont {Kimble}(2008)}]{kimble2008quantum}%
  \BibitemOpen
  \bibfield  {author} {\bibinfo {author} {\bibfnamefont {H.~J.}\ \bibnamefont
  {Kimble}},\ }\bibfield  {title} {\bibinfo {title} {The quantum internet},\
  }\href@noop {} {\bibfield  {journal} {\bibinfo  {journal} {Nature}\ }\textbf
  {\bibinfo {volume} {453}},\ \bibinfo {pages} {1023} (\bibinfo {year}
  {2008})}\BibitemShut {NoStop}%
\bibitem [{\citenamefont {Raymer}\ and\ \citenamefont
  {Srinivasan}(2012)}]{raymer2012manipulating}%
  \BibitemOpen
  \bibfield  {author} {\bibinfo {author} {\bibfnamefont {M.~G.}\ \bibnamefont
  {Raymer}}\ and\ \bibinfo {author} {\bibfnamefont {K.}~\bibnamefont
  {Srinivasan}},\ }\bibfield  {title} {\bibinfo {title} {Manipulating the color
  and shape of single photons},\ }\href@noop {} {\bibfield  {journal} {\bibinfo
   {journal} {Physics Today}\ }\textbf {\bibinfo {volume} {65}},\ \bibinfo
  {pages} {32} (\bibinfo {year} {2012})}\BibitemShut {NoStop}%
\bibitem [{\citenamefont {Rueda}\ \emph {et~al.}(2016)\citenamefont {Rueda},
  \citenamefont {Sedlmeir}, \citenamefont {Collodo}, \citenamefont {Vogl},
  \citenamefont {Stiller}, \citenamefont {Schunk}, \citenamefont {Strekalov},
  \citenamefont {Marquardt}, \citenamefont {Fink}, \citenamefont {Painter},
  \citenamefont {Leuchs},\ and\ \citenamefont
  {Schwefel}}]{rueda_efficient_2016}%
  \BibitemOpen
  \bibfield  {author} {\bibinfo {author} {\bibfnamefont {A.}~\bibnamefont
  {Rueda}}, \bibinfo {author} {\bibfnamefont {F.}~\bibnamefont {Sedlmeir}},
  \bibinfo {author} {\bibfnamefont {M.~C.}\ \bibnamefont {Collodo}}, \bibinfo
  {author} {\bibfnamefont {U.}~\bibnamefont {Vogl}}, \bibinfo {author}
  {\bibfnamefont {B.}~\bibnamefont {Stiller}}, \bibinfo {author} {\bibfnamefont
  {G.}~\bibnamefont {Schunk}}, \bibinfo {author} {\bibfnamefont {D.~V.}\
  \bibnamefont {Strekalov}}, \bibinfo {author} {\bibfnamefont {C.}~\bibnamefont
  {Marquardt}}, \bibinfo {author} {\bibfnamefont {J.~M.}\ \bibnamefont {Fink}},
  \bibinfo {author} {\bibfnamefont {O.}~\bibnamefont {Painter}}, \bibinfo
  {author} {\bibfnamefont {G.}~\bibnamefont {Leuchs}},\ and\ \bibinfo {author}
  {\bibfnamefont {H.~G.~L.}\ \bibnamefont {Schwefel}},\ }\bibfield  {title}
  {\bibinfo {title} {Efficient microwave to optical photon conversion: an
  electro-optical realization},\ }\href@noop {} {\bibfield  {journal} {\bibinfo
   {journal} {Optica}\ }\textbf {\bibinfo {volume} {3}},\ \bibinfo {pages}
  {597} (\bibinfo {year} {2016})}\BibitemShut {NoStop}%
\bibitem [{\citenamefont {Fan}\ \emph {et~al.}(2018)\citenamefont {Fan},
  \citenamefont {Zou}, \citenamefont {Cheng}, \citenamefont {Guo},
  \citenamefont {Han}, \citenamefont {Gong}, \citenamefont {Wang},\ and\
  \citenamefont {Tang}}]{fan_superconducting_2018}%
  \BibitemOpen
  \bibfield  {author} {\bibinfo {author} {\bibfnamefont {L.}~\bibnamefont
  {Fan}}, \bibinfo {author} {\bibfnamefont {C.-L.}\ \bibnamefont {Zou}},
  \bibinfo {author} {\bibfnamefont {R.}~\bibnamefont {Cheng}}, \bibinfo
  {author} {\bibfnamefont {X.}~\bibnamefont {Guo}}, \bibinfo {author}
  {\bibfnamefont {X.}~\bibnamefont {Han}}, \bibinfo {author} {\bibfnamefont
  {Z.}~\bibnamefont {Gong}}, \bibinfo {author} {\bibfnamefont {S.}~\bibnamefont
  {Wang}},\ and\ \bibinfo {author} {\bibfnamefont {H.~X.}\ \bibnamefont
  {Tang}},\ }\bibfield  {title} {\bibinfo {title} {Superconducting cavity
  electro-optics: {A} platform for coherent photon conversion between
  superconducting and photonic circuits},\ }\href@noop {} {\bibfield  {journal}
  {\bibinfo  {journal} {Sci. Adv.}\ }\textbf {\bibinfo {volume} {4}},\ \bibinfo
  {pages} {8} (\bibinfo {year} {2018})}\BibitemShut {NoStop}%
\bibitem [{\citenamefont {Hease}\ \emph {et~al.}(2020)\citenamefont {Hease},
  \citenamefont {Rueda}, \citenamefont {Sahu}, \citenamefont {Wulf},
  \citenamefont {Arnold}, \citenamefont {Schwefel},\ and\ \citenamefont
  {Fink}}]{hease2020bidirectional}%
  \BibitemOpen
  \bibfield  {author} {\bibinfo {author} {\bibfnamefont {W.}~\bibnamefont
  {Hease}}, \bibinfo {author} {\bibfnamefont {A.}~\bibnamefont {Rueda}},
  \bibinfo {author} {\bibfnamefont {R.}~\bibnamefont {Sahu}}, \bibinfo {author}
  {\bibfnamefont {M.}~\bibnamefont {Wulf}}, \bibinfo {author} {\bibfnamefont
  {G.}~\bibnamefont {Arnold}}, \bibinfo {author} {\bibfnamefont {H.~G.}\
  \bibnamefont {Schwefel}},\ and\ \bibinfo {author} {\bibfnamefont {J.~M.}\
  \bibnamefont {Fink}},\ }\bibfield  {title} {\bibinfo {title} {Bidirectional
  electro-optic wavelength conversion in the quantum ground state},\
  }\href@noop {} {\bibfield  {journal} {\bibinfo  {journal} {PRX Quantum}\
  }\textbf {\bibinfo {volume} {1}},\ \bibinfo {pages} {020315} (\bibinfo {year}
  {2020})}\BibitemShut {NoStop}%
\bibitem [{\citenamefont {Fu}\ \emph {et~al.}(2020)\citenamefont {Fu},
  \citenamefont {Xu}, \citenamefont {Liu}, \citenamefont {Zou}, \citenamefont
  {Zhong}, \citenamefont {Han}, \citenamefont {Shen}, \citenamefont {Xu},
  \citenamefont {Cheng}, \citenamefont {Wang}, \citenamefont {Jiang},\ and\
  \citenamefont {Tang}}]{fu_ground-state_2020}%
  \BibitemOpen
  \bibfield  {author} {\bibinfo {author} {\bibfnamefont {W.}~\bibnamefont
  {Fu}}, \bibinfo {author} {\bibfnamefont {M.}~\bibnamefont {Xu}}, \bibinfo
  {author} {\bibfnamefont {X.}~\bibnamefont {Liu}}, \bibinfo {author}
  {\bibfnamefont {C.-L.}\ \bibnamefont {Zou}}, \bibinfo {author} {\bibfnamefont
  {C.}~\bibnamefont {Zhong}}, \bibinfo {author} {\bibfnamefont
  {X.}~\bibnamefont {Han}}, \bibinfo {author} {\bibfnamefont {M.}~\bibnamefont
  {Shen}}, \bibinfo {author} {\bibfnamefont {Y.}~\bibnamefont {Xu}}, \bibinfo
  {author} {\bibfnamefont {R.}~\bibnamefont {Cheng}}, \bibinfo {author}
  {\bibfnamefont {S.}~\bibnamefont {Wang}}, \bibinfo {author} {\bibfnamefont
  {L.}~\bibnamefont {Jiang}},\ and\ \bibinfo {author} {\bibfnamefont {H.~X.}\
  \bibnamefont {Tang}},\ }\bibfield  {title} {\bibinfo {title} {Ground-state
  {Pulsed} {Cavity} {Electro}-optics for {Microwave}-to-optical {Conversion}},\
  }\href {http://arxiv.org/abs/2010.11392} {\bibfield  {journal} {\bibinfo
  {journal} {arXiv:2010.11392 [physics, physics:quant-ph]}\ } (\bibinfo {year}
  {2020})},\ \bibinfo {note} {arXiv: 2010.11392}\BibitemShut {NoStop}%
\bibitem [{\citenamefont {Bochmann}\ \emph {et~al.}(2013)\citenamefont
  {Bochmann}, \citenamefont {Vainsencher}, \citenamefont {Awschalom},\ and\
  \citenamefont {Cleland}}]{bochmann_nanomechanical_2013}%
  \BibitemOpen
  \bibfield  {author} {\bibinfo {author} {\bibfnamefont {J.}~\bibnamefont
  {Bochmann}}, \bibinfo {author} {\bibfnamefont {A.}~\bibnamefont
  {Vainsencher}}, \bibinfo {author} {\bibfnamefont {D.~D.}\ \bibnamefont
  {Awschalom}},\ and\ \bibinfo {author} {\bibfnamefont {A.~N.}\ \bibnamefont
  {Cleland}},\ }\bibfield  {title} {\bibinfo {title} {Nanomechanical coupling
  between microwave and optical photons},\ }\href
  {https://doi.org/10.1038/nphys2748} {\bibfield  {journal} {\bibinfo
  {journal} {Nat. Phys.}\ }\textbf {\bibinfo {volume} {9}},\ \bibinfo {pages}
  {712} (\bibinfo {year} {2013})}\BibitemShut {NoStop}%
\bibitem [{\citenamefont {Balram}\ \emph {et~al.}(2016)\citenamefont {Balram},
  \citenamefont {Davan{\c{c}}o}, \citenamefont {Song},\ and\ \citenamefont
  {Srinivasan}}]{balram_coherent_2016}%
  \BibitemOpen
  \bibfield  {author} {\bibinfo {author} {\bibfnamefont {K.~C.}\ \bibnamefont
  {Balram}}, \bibinfo {author} {\bibfnamefont {M.~I.}\ \bibnamefont
  {Davan{\c{c}}o}}, \bibinfo {author} {\bibfnamefont {J.~D.}\ \bibnamefont
  {Song}},\ and\ \bibinfo {author} {\bibfnamefont {K.}~\bibnamefont
  {Srinivasan}},\ }\bibfield  {title} {\bibinfo {title} {Coherent coupling
  between radiofrequency, optical and acoustic waves in piezo-optomechanical
  circuits},\ }\href {https://doi.org/10.1038/nphoton.2016.46} {\bibfield
  {journal} {\bibinfo  {journal} {Nat. Photonics}\ }\textbf {\bibinfo {volume}
  {10}},\ \bibinfo {pages} {346} (\bibinfo {year} {2016})}\BibitemShut
  {NoStop}%
\bibitem [{\citenamefont {Jiang}\ \emph {et~al.}(2020)\citenamefont {Jiang},
  \citenamefont {Sarabalis}, \citenamefont {Dahmani}, \citenamefont {Patel},
  \citenamefont {Mayor}, \citenamefont {McKenna}, \citenamefont {Van~Laer},\
  and\ \citenamefont {Safavi-Naeini}}]{jiang_efficient_2020}%
  \BibitemOpen
  \bibfield  {author} {\bibinfo {author} {\bibfnamefont {W.}~\bibnamefont
  {Jiang}}, \bibinfo {author} {\bibfnamefont {C.~J.}\ \bibnamefont
  {Sarabalis}}, \bibinfo {author} {\bibfnamefont {Y.~D.}\ \bibnamefont
  {Dahmani}}, \bibinfo {author} {\bibfnamefont {R.~N.}\ \bibnamefont {Patel}},
  \bibinfo {author} {\bibfnamefont {F.~M.}\ \bibnamefont {Mayor}}, \bibinfo
  {author} {\bibfnamefont {T.~P.}\ \bibnamefont {McKenna}}, \bibinfo {author}
  {\bibfnamefont {R.}~\bibnamefont {Van~Laer}},\ and\ \bibinfo {author}
  {\bibfnamefont {A.~H.}\ \bibnamefont {Safavi-Naeini}},\ }\bibfield  {title}
  {\bibinfo {title} {Efficient bidirectional piezo-optomechanical transduction
  between microwave and optical frequency},\ }\href
  {https://doi.org/10.1038/s41467-020-14863-3} {\bibfield  {journal} {\bibinfo
  {journal} {Nature Communications}\ }\textbf {\bibinfo {volume} {11}},\
  \bibinfo {pages} {1166} (\bibinfo {year} {2020})}\BibitemShut {NoStop}%
\bibitem [{\citenamefont {Han}\ \emph {et~al.}(2020)\citenamefont {Han},
  \citenamefont {Fu}, \citenamefont {Zhong}, \citenamefont {Zou}, \citenamefont
  {Xu}, \citenamefont {Sayem}, \citenamefont {Xu}, \citenamefont {Wang},
  \citenamefont {Cheng}, \citenamefont {Jiang},\ and\ \citenamefont
  {Tang}}]{han_cavity_2020}%
  \BibitemOpen
  \bibfield  {author} {\bibinfo {author} {\bibfnamefont {X.}~\bibnamefont
  {Han}}, \bibinfo {author} {\bibfnamefont {W.}~\bibnamefont {Fu}}, \bibinfo
  {author} {\bibfnamefont {C.}~\bibnamefont {Zhong}}, \bibinfo {author}
  {\bibfnamefont {C.-L.}\ \bibnamefont {Zou}}, \bibinfo {author} {\bibfnamefont
  {Y.}~\bibnamefont {Xu}}, \bibinfo {author} {\bibfnamefont {A.~A.}\
  \bibnamefont {Sayem}}, \bibinfo {author} {\bibfnamefont {M.}~\bibnamefont
  {Xu}}, \bibinfo {author} {\bibfnamefont {S.}~\bibnamefont {Wang}}, \bibinfo
  {author} {\bibfnamefont {R.}~\bibnamefont {Cheng}}, \bibinfo {author}
  {\bibfnamefont {L.}~\bibnamefont {Jiang}},\ and\ \bibinfo {author}
  {\bibfnamefont {H.~X.}\ \bibnamefont {Tang}},\ }\bibfield  {title} {\bibinfo
  {title} {Cavity piezo-mechanics for superconducting-nanophotonic quantum
  interface},\ }\href {https://doi.org/10.1038/s41467-020-17053-3} {\bibfield
  {journal} {\bibinfo  {journal} {Nature Communications}\ }\textbf {\bibinfo
  {volume} {11}},\ \bibinfo {pages} {3237} (\bibinfo {year}
  {2020})}\BibitemShut {NoStop}%
\bibitem [{\citenamefont {Mirhosseini}\ \emph {et~al.}(2020)\citenamefont
  {Mirhosseini}, \citenamefont {Sipahigil}, \citenamefont {Kalaee},\ and\
  \citenamefont {Painter}}]{mirhosseini2020superconducting}%
  \BibitemOpen
  \bibfield  {author} {\bibinfo {author} {\bibfnamefont {M.}~\bibnamefont
  {Mirhosseini}}, \bibinfo {author} {\bibfnamefont {A.}~\bibnamefont
  {Sipahigil}}, \bibinfo {author} {\bibfnamefont {M.}~\bibnamefont {Kalaee}},\
  and\ \bibinfo {author} {\bibfnamefont {O.}~\bibnamefont {Painter}},\
  }\bibfield  {title} {\bibinfo {title} {Superconducting qubit to optical
  photon transduction},\ }\href@noop {} {\bibfield  {journal} {\bibinfo
  {journal} {Nature}\ }\textbf {\bibinfo {volume} {588}},\ \bibinfo {pages}
  {599} (\bibinfo {year} {2020})}\BibitemShut {NoStop}%
\bibitem [{\citenamefont {Aspelmeyer}\ \emph {et~al.}(2014)\citenamefont
  {Aspelmeyer}, \citenamefont {Kippenberg},\ and\ \citenamefont
  {Marquardt}}]{aspelmeyer_cavity_2014}%
  \BibitemOpen
  \bibfield  {author} {\bibinfo {author} {\bibfnamefont {M.}~\bibnamefont
  {Aspelmeyer}}, \bibinfo {author} {\bibfnamefont {T.~J.}\ \bibnamefont
  {Kippenberg}},\ and\ \bibinfo {author} {\bibfnamefont {F.}~\bibnamefont
  {Marquardt}},\ }\bibfield  {title} {\bibinfo {title} {Cavity optomechanics},\
  }\href {https://doi.org/10.1103/RevModPhys.86.1391} {\bibfield  {journal}
  {\bibinfo  {journal} {Rev. Mod. Phys.}\ }\textbf {\bibinfo {volume} {86}},\
  \bibinfo {pages} {1391} (\bibinfo {year} {2014})}\BibitemShut {NoStop}%
\bibitem [{\citenamefont {Safavi-Naeini}\ \emph {et~al.}(2019)\citenamefont
  {Safavi-Naeini}, \citenamefont {Van~Thourhout}, \citenamefont {Baets},\ and\
  \citenamefont {Van~Laer}}]{safavi2019controlling}%
  \BibitemOpen
  \bibfield  {author} {\bibinfo {author} {\bibfnamefont {A.~H.}\ \bibnamefont
  {Safavi-Naeini}}, \bibinfo {author} {\bibfnamefont {D.}~\bibnamefont
  {Van~Thourhout}}, \bibinfo {author} {\bibfnamefont {R.}~\bibnamefont
  {Baets}},\ and\ \bibinfo {author} {\bibfnamefont {R.}~\bibnamefont
  {Van~Laer}},\ }\bibfield  {title} {\bibinfo {title} {Controlling phonons and
  photons at the wavelength scale: integrated photonics meets integrated
  phononics},\ }\href@noop {} {\bibfield  {journal} {\bibinfo  {journal}
  {Optica}\ }\textbf {\bibinfo {volume} {6}},\ \bibinfo {pages} {213} (\bibinfo
  {year} {2019})}\BibitemShut {NoStop}%
\bibitem [{\citenamefont {O’Connell}\ \emph {et~al.}(2010)\citenamefont
  {O’Connell}, \citenamefont {Hofheinz}, \citenamefont {Ansmann},
  \citenamefont {Bialczak}, \citenamefont {Lenander}, \citenamefont {Lucero},
  \citenamefont {Neeley}, \citenamefont {Sank}, \citenamefont {Wang},
  \citenamefont {Weides} \emph {et~al.}}]{o2010quantum}%
  \BibitemOpen
  \bibfield  {author} {\bibinfo {author} {\bibfnamefont {A.~D.}\ \bibnamefont
  {O’Connell}}, \bibinfo {author} {\bibfnamefont {M.}~\bibnamefont
  {Hofheinz}}, \bibinfo {author} {\bibfnamefont {M.}~\bibnamefont {Ansmann}},
  \bibinfo {author} {\bibfnamefont {R.~C.}\ \bibnamefont {Bialczak}}, \bibinfo
  {author} {\bibfnamefont {M.}~\bibnamefont {Lenander}}, \bibinfo {author}
  {\bibfnamefont {E.}~\bibnamefont {Lucero}}, \bibinfo {author} {\bibfnamefont
  {M.}~\bibnamefont {Neeley}}, \bibinfo {author} {\bibfnamefont
  {D.}~\bibnamefont {Sank}}, \bibinfo {author} {\bibfnamefont {H.}~\bibnamefont
  {Wang}}, \bibinfo {author} {\bibfnamefont {M.}~\bibnamefont {Weides}}, \emph
  {et~al.},\ }\bibfield  {title} {\bibinfo {title} {Quantum ground state and
  single-phonon control of a mechanical resonator},\ }\href@noop {} {\bibfield
  {journal} {\bibinfo  {journal} {Nature}\ }\textbf {\bibinfo {volume} {464}},\
  \bibinfo {pages} {697} (\bibinfo {year} {2010})}\BibitemShut {NoStop}%
\bibitem [{\citenamefont {Satzinger}\ \emph {et~al.}(2018)\citenamefont
  {Satzinger}, \citenamefont {Zhong}, \citenamefont {Chang}, \citenamefont
  {Peairs}, \citenamefont {Bienfait}, \citenamefont {Chou}, \citenamefont
  {Cleland}, \citenamefont {Conner}, \citenamefont {Dumur}, \citenamefont
  {Grebel} \emph {et~al.}}]{satzinger2018quantum}%
  \BibitemOpen
  \bibfield  {author} {\bibinfo {author} {\bibfnamefont {K.~J.}\ \bibnamefont
  {Satzinger}}, \bibinfo {author} {\bibfnamefont {Y.}~\bibnamefont {Zhong}},
  \bibinfo {author} {\bibfnamefont {H.-S.}\ \bibnamefont {Chang}}, \bibinfo
  {author} {\bibfnamefont {G.~A.}\ \bibnamefont {Peairs}}, \bibinfo {author}
  {\bibfnamefont {A.}~\bibnamefont {Bienfait}}, \bibinfo {author}
  {\bibfnamefont {M.-H.}\ \bibnamefont {Chou}}, \bibinfo {author}
  {\bibfnamefont {A.}~\bibnamefont {Cleland}}, \bibinfo {author} {\bibfnamefont
  {C.~R.}\ \bibnamefont {Conner}}, \bibinfo {author} {\bibfnamefont
  {{\'E}.}~\bibnamefont {Dumur}}, \bibinfo {author} {\bibfnamefont
  {J.}~\bibnamefont {Grebel}}, \emph {et~al.},\ }\bibfield  {title} {\bibinfo
  {title} {Quantum control of surface acoustic-wave phonons},\ }\href@noop {}
  {\bibfield  {journal} {\bibinfo  {journal} {Nature}\ }\textbf {\bibinfo
  {volume} {563}},\ \bibinfo {pages} {661} (\bibinfo {year}
  {2018})}\BibitemShut {NoStop}%
\bibitem [{\citenamefont {Chu}\ \emph {et~al.}(2017)\citenamefont {Chu},
  \citenamefont {Kharel}, \citenamefont {Renninger}, \citenamefont {Burkhart},
  \citenamefont {Frunzio}, \citenamefont {Rakich},\ and\ \citenamefont
  {Schoelkopf}}]{chu2017quantum}%
  \BibitemOpen
  \bibfield  {author} {\bibinfo {author} {\bibfnamefont {Y.}~\bibnamefont
  {Chu}}, \bibinfo {author} {\bibfnamefont {P.}~\bibnamefont {Kharel}},
  \bibinfo {author} {\bibfnamefont {W.~H.}\ \bibnamefont {Renninger}}, \bibinfo
  {author} {\bibfnamefont {L.~D.}\ \bibnamefont {Burkhart}}, \bibinfo {author}
  {\bibfnamefont {L.}~\bibnamefont {Frunzio}}, \bibinfo {author} {\bibfnamefont
  {P.~T.}\ \bibnamefont {Rakich}},\ and\ \bibinfo {author} {\bibfnamefont
  {R.~J.}\ \bibnamefont {Schoelkopf}},\ }\bibfield  {title} {\bibinfo {title}
  {Quantum acoustics with superconducting qubits},\ }\href@noop {} {\bibfield
  {journal} {\bibinfo  {journal} {Science}\ }\textbf {\bibinfo {volume}
  {358}},\ \bibinfo {pages} {199} (\bibinfo {year} {2017})}\BibitemShut
  {NoStop}%
\bibitem [{\citenamefont {Moores}\ \emph {et~al.}(2018)\citenamefont {Moores},
  \citenamefont {Sletten}, \citenamefont {Viennot},\ and\ \citenamefont
  {Lehnert}}]{moores2018cavity}%
  \BibitemOpen
  \bibfield  {author} {\bibinfo {author} {\bibfnamefont {B.~A.}\ \bibnamefont
  {Moores}}, \bibinfo {author} {\bibfnamefont {L.~R.}\ \bibnamefont {Sletten}},
  \bibinfo {author} {\bibfnamefont {J.~J.}\ \bibnamefont {Viennot}},\ and\
  \bibinfo {author} {\bibfnamefont {K.}~\bibnamefont {Lehnert}},\ }\bibfield
  {title} {\bibinfo {title} {Cavity quantum acoustic device in the multimode
  strong coupling regime},\ }\href@noop {} {\bibfield  {journal} {\bibinfo
  {journal} {Physical review letters}\ }\textbf {\bibinfo {volume} {120}},\
  \bibinfo {pages} {227701} (\bibinfo {year} {2018})}\BibitemShut {NoStop}%
\bibitem [{\citenamefont {MacCabe}\ \emph
  {et~al.}(2020{\natexlab{a}})\citenamefont {MacCabe}, \citenamefont {Ren},
  \citenamefont {Luo}, \citenamefont {Cohen}, \citenamefont {Zhou},
  \citenamefont {Sipahigil}, \citenamefont {Mirhosseini},\ and\ \citenamefont
  {Painter}}]{maccabe2020nano}%
  \BibitemOpen
  \bibfield  {author} {\bibinfo {author} {\bibfnamefont {G.~S.}\ \bibnamefont
  {MacCabe}}, \bibinfo {author} {\bibfnamefont {H.}~\bibnamefont {Ren}},
  \bibinfo {author} {\bibfnamefont {J.}~\bibnamefont {Luo}}, \bibinfo {author}
  {\bibfnamefont {J.~D.}\ \bibnamefont {Cohen}}, \bibinfo {author}
  {\bibfnamefont {H.}~\bibnamefont {Zhou}}, \bibinfo {author} {\bibfnamefont
  {A.}~\bibnamefont {Sipahigil}}, \bibinfo {author} {\bibfnamefont
  {M.}~\bibnamefont {Mirhosseini}},\ and\ \bibinfo {author} {\bibfnamefont
  {O.}~\bibnamefont {Painter}},\ }\bibfield  {title} {\bibinfo {title}
  {Nano-acoustic resonator with ultralong phonon lifetime},\ }\href@noop {}
  {\bibfield  {journal} {\bibinfo  {journal} {Science}\ }\textbf {\bibinfo
  {volume} {370}},\ \bibinfo {pages} {840} (\bibinfo {year}
  {2020}{\natexlab{a}})}\BibitemShut {NoStop}%
\bibitem [{\citenamefont {Gokhale}\ \emph {et~al.}(2020)\citenamefont
  {Gokhale}, \citenamefont {Downey}, \citenamefont {Katzer}, \citenamefont
  {Nepal}, \citenamefont {Lang}, \citenamefont {Stroud},\ and\ \citenamefont
  {Meyer}}]{gokhale2020epitaxial}%
  \BibitemOpen
  \bibfield  {author} {\bibinfo {author} {\bibfnamefont {V.~J.}\ \bibnamefont
  {Gokhale}}, \bibinfo {author} {\bibfnamefont {B.~P.}\ \bibnamefont {Downey}},
  \bibinfo {author} {\bibfnamefont {D.~S.}\ \bibnamefont {Katzer}}, \bibinfo
  {author} {\bibfnamefont {N.}~\bibnamefont {Nepal}}, \bibinfo {author}
  {\bibfnamefont {A.~C.}\ \bibnamefont {Lang}}, \bibinfo {author}
  {\bibfnamefont {R.~M.}\ \bibnamefont {Stroud}},\ and\ \bibinfo {author}
  {\bibfnamefont {D.~J.}\ \bibnamefont {Meyer}},\ }\bibfield  {title} {\bibinfo
  {title} {Epitaxial bulk acoustic wave resonators as highly coherent
  multi-phonon sources for quantum acoustodynamics},\ }\href@noop {} {\bibfield
   {journal} {\bibinfo  {journal} {Nature communications}\ }\textbf {\bibinfo
  {volume} {11}},\ \bibinfo {pages} {1} (\bibinfo {year} {2020})}\BibitemShut
  {NoStop}%
\bibitem [{\citenamefont {Kervinen}\ \emph {et~al.}(2018)\citenamefont
  {Kervinen}, \citenamefont {Rissanen},\ and\ \citenamefont
  {Sillanp{\"a}{\"a}}}]{kervinen2018interfacing}%
  \BibitemOpen
  \bibfield  {author} {\bibinfo {author} {\bibfnamefont {M.}~\bibnamefont
  {Kervinen}}, \bibinfo {author} {\bibfnamefont {I.}~\bibnamefont {Rissanen}},\
  and\ \bibinfo {author} {\bibfnamefont {M.}~\bibnamefont
  {Sillanp{\"a}{\"a}}},\ }\bibfield  {title} {\bibinfo {title} {Interfacing
  planar superconducting qubits with high overtone bulk acoustic phonons},\
  }\href@noop {} {\bibfield  {journal} {\bibinfo  {journal} {Physical Review
  B}\ }\textbf {\bibinfo {volume} {97}},\ \bibinfo {pages} {205443} (\bibinfo
  {year} {2018})}\BibitemShut {NoStop}%
\bibitem [{\citenamefont {Forsch}\ \emph
  {et~al.}(2020{\natexlab{a}})\citenamefont {Forsch}, \citenamefont {Stockill},
  \citenamefont {Wallucks}, \citenamefont {Marinkovic}, \citenamefont
  {Gärtner}, \citenamefont {Norte}, \citenamefont {van Otten}, \citenamefont
  {Fiore}, \citenamefont {Srinivasan},\ and\ \citenamefont
  {Groeblacher}}]{forsch_microwave_2018}%
  \BibitemOpen
  \bibfield  {author} {\bibinfo {author} {\bibfnamefont {M.}~\bibnamefont
  {Forsch}}, \bibinfo {author} {\bibfnamefont {R.}~\bibnamefont {Stockill}},
  \bibinfo {author} {\bibfnamefont {A.}~\bibnamefont {Wallucks}}, \bibinfo
  {author} {\bibfnamefont {I.}~\bibnamefont {Marinkovic}}, \bibinfo {author}
  {\bibfnamefont {C.}~\bibnamefont {Gärtner}}, \bibinfo {author}
  {\bibfnamefont {R.~A.}\ \bibnamefont {Norte}}, \bibinfo {author}
  {\bibfnamefont {F.}~\bibnamefont {van Otten}}, \bibinfo {author}
  {\bibfnamefont {A.}~\bibnamefont {Fiore}}, \bibinfo {author} {\bibfnamefont
  {K.}~\bibnamefont {Srinivasan}},\ and\ \bibinfo {author} {\bibfnamefont
  {S.}~\bibnamefont {Groeblacher}},\ }\bibfield  {title} {\bibinfo {title}
  {Microwave-to-optics conversion using a mechanical oscillator in its quantum
  ground state},\ }\href@noop {} {\bibfield  {journal} {\bibinfo  {journal}
  {Nat. Phys}\ }\textbf {\bibinfo {volume} {16}},\ \bibinfo {pages} {69}
  (\bibinfo {year} {2020}{\natexlab{a}})}\BibitemShut {NoStop}%
\bibitem [{\citenamefont {Tian}(2012)}]{Tian_adiabatic-state-conversion_2012}%
  \BibitemOpen
  \bibfield  {author} {\bibinfo {author} {\bibfnamefont {L.}~\bibnamefont
  {Tian}},\ }\bibfield  {title} {\bibinfo {title} {Adiabatic state conversion
  and pulse transmission in optomechanical systems},\ }\href
  {https://doi.org/10.1103/PhysRevLett.108.153604} {\bibfield  {journal}
  {\bibinfo  {journal} {Phys. Rev. Lett.}\ }\textbf {\bibinfo {volume} {108}},\
  \bibinfo {pages} {153604} (\bibinfo {year} {2012})}\BibitemShut {NoStop}%
\bibitem [{\citenamefont {Wang}\ and\ \citenamefont
  {Clerk}(2012)}]{Wang_quantum-state-transfer_2012}%
  \BibitemOpen
  \bibfield  {author} {\bibinfo {author} {\bibfnamefont {Y.-D.}\ \bibnamefont
  {Wang}}\ and\ \bibinfo {author} {\bibfnamefont {A.~A.}\ \bibnamefont
  {Clerk}},\ }\bibfield  {title} {\bibinfo {title} {Using interference for high
  fidelity quantum state transfer in optomechanics},\ }\href
  {https://doi.org/10.1103/PhysRevLett.108.153603} {\bibfield  {journal}
  {\bibinfo  {journal} {Phys. Rev. Lett.}\ }\textbf {\bibinfo {volume} {108}},\
  \bibinfo {pages} {153603} (\bibinfo {year} {2012})}\BibitemShut {NoStop}%
\bibitem [{\citenamefont {Andrews}\ \emph {et~al.}(2014)\citenamefont
  {Andrews}, \citenamefont {Peterson}, \citenamefont {Purdy}, \citenamefont
  {Cicak}, \citenamefont {Simmonds}, \citenamefont {Regal},\ and\ \citenamefont
  {Lehnert}}]{Andrews2014}%
  \BibitemOpen
  \bibfield  {author} {\bibinfo {author} {\bibfnamefont {R.~W.}\ \bibnamefont
  {Andrews}}, \bibinfo {author} {\bibfnamefont {R.~W.}\ \bibnamefont
  {Peterson}}, \bibinfo {author} {\bibfnamefont {T.~P.}\ \bibnamefont {Purdy}},
  \bibinfo {author} {\bibfnamefont {K.}~\bibnamefont {Cicak}}, \bibinfo
  {author} {\bibfnamefont {R.~W.}\ \bibnamefont {Simmonds}}, \bibinfo {author}
  {\bibfnamefont {C.~A.}\ \bibnamefont {Regal}},\ and\ \bibinfo {author}
  {\bibfnamefont {K.~W.}\ \bibnamefont {Lehnert}},\ }\bibfield  {title}
  {\bibinfo {title} {Bidirectional and efficient conversion between microwave
  and optical light},\ }\href {http://dx.doi.org/10.1038/nphys2911} {\bibfield
  {journal} {\bibinfo  {journal} {Nat. Phys.}\ }\textbf {\bibinfo {volume}
  {10}},\ \bibinfo {pages} {321} (\bibinfo {year} {2014})}\BibitemShut
  {NoStop}%
\bibitem [{\citenamefont {Wu}\ \emph {et~al.}(2020)\citenamefont {Wu},
  \citenamefont {Zeuthen}, \citenamefont {Balram},\ and\ \citenamefont
  {Srinivasan}}]{wu2020microwave}%
  \BibitemOpen
  \bibfield  {author} {\bibinfo {author} {\bibfnamefont {M.}~\bibnamefont
  {Wu}}, \bibinfo {author} {\bibfnamefont {E.}~\bibnamefont {Zeuthen}},
  \bibinfo {author} {\bibfnamefont {K.~C.}\ \bibnamefont {Balram}},\ and\
  \bibinfo {author} {\bibfnamefont {K.}~\bibnamefont {Srinivasan}},\ }\bibfield
   {title} {\bibinfo {title} {Microwave-to-optical transduction using a
  mechanical supermode for coupling piezoelectric and optomechanical
  resonators},\ }\href@noop {} {\bibfield  {journal} {\bibinfo  {journal}
  {Physical Review Applied}\ }\textbf {\bibinfo {volume} {13}},\ \bibinfo
  {pages} {014027} (\bibinfo {year} {2020})}\BibitemShut {NoStop}%
\bibitem [{\citenamefont {Hashimoto}(2009)}]{hashimoto2009rf}%
  \BibitemOpen
  \bibfield  {author} {\bibinfo {author} {\bibfnamefont {K.-y.}\ \bibnamefont
  {Hashimoto}},\ }\href@noop {} {\emph {\bibinfo {title} {RF bulk acoustic wave
  filters for communications}}}\ (\bibinfo  {publisher} {Artech House},\
  \bibinfo {year} {2009})\BibitemShut {NoStop}%
\bibitem [{\citenamefont {Balram}\ \emph {et~al.}(2014)\citenamefont {Balram},
  \citenamefont {Davan{\c{c}}o}, \citenamefont {Lim}, \citenamefont {Song},\
  and\ \citenamefont {Srinivasan}}]{Balram_Moving_2014}%
  \BibitemOpen
  \bibfield  {author} {\bibinfo {author} {\bibfnamefont {K.~C.}\ \bibnamefont
  {Balram}}, \bibinfo {author} {\bibfnamefont {M.}~\bibnamefont
  {Davan{\c{c}}o}}, \bibinfo {author} {\bibfnamefont {J.~Y.}\ \bibnamefont
  {Lim}}, \bibinfo {author} {\bibfnamefont {J.~D.}\ \bibnamefont {Song}},\ and\
  \bibinfo {author} {\bibfnamefont {K.}~\bibnamefont {Srinivasan}},\ }\bibfield
   {title} {\bibinfo {title} {{Moving boundary and photoelastic coupling in
  GaAs optomechanical resonators}},\ }\href
  {https://doi.org/10.1364/OPTICA.1.000414} {\bibfield  {journal} {\bibinfo
  {journal} {Optica}\ }\textbf {\bibinfo {volume} {1}},\ \bibinfo {pages} {414}
  (\bibinfo {year} {2014})}\BibitemShut {NoStop}%
\bibitem [{\citenamefont {Van~Laer}\ \emph {et~al.}(2016)\citenamefont
  {Van~Laer}, \citenamefont {Baets},\ and\ \citenamefont
  {Van~Thourhout}}]{van2016unifying}%
  \BibitemOpen
  \bibfield  {author} {\bibinfo {author} {\bibfnamefont {R.}~\bibnamefont
  {Van~Laer}}, \bibinfo {author} {\bibfnamefont {R.}~\bibnamefont {Baets}},\
  and\ \bibinfo {author} {\bibfnamefont {D.}~\bibnamefont {Van~Thourhout}},\
  }\bibfield  {title} {\bibinfo {title} {Unifying brillouin scattering and
  cavity optomechanics},\ }\href@noop {} {\bibfield  {journal} {\bibinfo
  {journal} {Physical Review A}\ }\textbf {\bibinfo {volume} {93}},\ \bibinfo
  {pages} {053828} (\bibinfo {year} {2016})}\BibitemShut {NoStop}%
\bibitem [{\citenamefont {Renninger}\ \emph {et~al.}(2018)\citenamefont
  {Renninger}, \citenamefont {Kharel}, \citenamefont {Behunin},\ and\
  \citenamefont {Rakich}}]{renninger2018bulk}%
  \BibitemOpen
  \bibfield  {author} {\bibinfo {author} {\bibfnamefont {W.}~\bibnamefont
  {Renninger}}, \bibinfo {author} {\bibfnamefont {P.}~\bibnamefont {Kharel}},
  \bibinfo {author} {\bibfnamefont {R.}~\bibnamefont {Behunin}},\ and\ \bibinfo
  {author} {\bibfnamefont {P.}~\bibnamefont {Rakich}},\ }\bibfield  {title}
  {\bibinfo {title} {Bulk crystalline optomechanics},\ }\href@noop {}
  {\bibfield  {journal} {\bibinfo  {journal} {Nature Physics}\ }\textbf
  {\bibinfo {volume} {14}},\ \bibinfo {pages} {601} (\bibinfo {year}
  {2018})}\BibitemShut {NoStop}%
\bibitem [{\citenamefont {Valle}\ and\ \citenamefont
  {Balram}(2019)}]{valle_high-frequency_2019}%
  \BibitemOpen
  \bibfield  {author} {\bibinfo {author} {\bibfnamefont {S.}~\bibnamefont
  {Valle}}\ and\ \bibinfo {author} {\bibfnamefont {K.~C.}\ \bibnamefont
  {Balram}},\ }\bibfield  {title} {\bibinfo {title} {High-frequency, resonant
  acousto-optic modulators fabricated in a {MEMS} foundry platform},\
  }\href@noop {} {\bibfield  {journal} {\bibinfo  {journal} {Opt. Lett.}\
  }\textbf {\bibinfo {volume} {44}},\ \bibinfo {pages} {3777} (\bibinfo {year}
  {2019})}\BibitemShut {NoStop}%
\bibitem [{\citenamefont {Campbell}(1998)}]{campbell_surface_1998}%
  \BibitemOpen
  \bibfield  {author} {\bibinfo {author} {\bibfnamefont {C.}~\bibnamefont
  {Campbell}},\ }\href@noop {} {\emph {\bibinfo {title} {Surface acoustic wave
  devices for mobile and wireless communications}}}\ (\bibinfo  {publisher}
  {Academic press},\ \bibinfo {year} {1998})\BibitemShut {NoStop}%
\bibitem [{\citenamefont {Vainsencher}\ \emph {et~al.}(2016)\citenamefont
  {Vainsencher}, \citenamefont {Satzinger}, \citenamefont {Peairs},\ and\
  \citenamefont {Cleland}}]{Vainsencher_Bi_2016}%
  \BibitemOpen
  \bibfield  {author} {\bibinfo {author} {\bibfnamefont {A.}~\bibnamefont
  {Vainsencher}}, \bibinfo {author} {\bibfnamefont {K.~J.}\ \bibnamefont
  {Satzinger}}, \bibinfo {author} {\bibfnamefont {G.~A.}\ \bibnamefont
  {Peairs}},\ and\ \bibinfo {author} {\bibfnamefont {A.~N.}\ \bibnamefont
  {Cleland}},\ }\bibfield  {title} {\bibinfo {title} {Bi-directional conversion
  between microwave and optical frequencies in a piezoelectric optomechanical
  device},\ }\href@noop {} {\bibfield  {journal} {\bibinfo  {journal} {Appl.
  Phys. Lett.}\ }\textbf {\bibinfo {volume} {109}},\ \bibinfo {pages} {033107}
  (\bibinfo {year} {2016})}\BibitemShut {NoStop}%
\bibitem [{\citenamefont {Gong}\ and\ \citenamefont
  {Piazza}(2013)}]{gong_design_2013}%
  \BibitemOpen
  \bibfield  {author} {\bibinfo {author} {\bibfnamefont {S.}~\bibnamefont
  {Gong}}\ and\ \bibinfo {author} {\bibfnamefont {G.}~\bibnamefont {Piazza}},\
  }\bibfield  {title} {\bibinfo {title} {Design and analysis of lithium
  niobate-based high electromechanical coupling {RF}-{MEMS} resonators for
  wideband filtering},\ }\href {https://doi.org/10.1109/TMTT.2012.2228671}
  {\bibfield  {journal} {\bibinfo  {journal} {IEEE Trans. Microw. Theory and
  Tech.}\ }\textbf {\bibinfo {volume} {61}},\ \bibinfo {pages} {1} (\bibinfo
  {year} {2013})}\BibitemShut {NoStop}%
\bibitem [{\citenamefont {Chan}\ \emph {et~al.}(2012)\citenamefont {Chan},
  \citenamefont {Safavi-Naeini}, \citenamefont {Hill}, \citenamefont
  {Meenehan},\ and\ \citenamefont {Painter}}]{chan_optimized_2012}%
  \BibitemOpen
  \bibfield  {author} {\bibinfo {author} {\bibfnamefont {J.}~\bibnamefont
  {Chan}}, \bibinfo {author} {\bibfnamefont {A.~H.}\ \bibnamefont
  {Safavi-Naeini}}, \bibinfo {author} {\bibfnamefont {J.~T.}\ \bibnamefont
  {Hill}}, \bibinfo {author} {\bibfnamefont {S.}~\bibnamefont {Meenehan}},\
  and\ \bibinfo {author} {\bibfnamefont {O.}~\bibnamefont {Painter}},\
  }\bibfield  {title} {\bibinfo {title} {Optimized optomechanical crystal
  cavity with acoustic radiation shield},\ }\href
  {https://doi.org/10.1063/1.4747726} {\bibfield  {journal} {\bibinfo
  {journal} {Appl. Phys. Lett.}\ }\textbf {\bibinfo {volume} {101}},\ \bibinfo
  {pages} {081115} (\bibinfo {year} {2012})}\BibitemShut {NoStop}%
\bibitem [{\citenamefont {Hunger}\ \emph {et~al.}(2010)\citenamefont {Hunger},
  \citenamefont {Steinmetz}, \citenamefont {Colombe}, \citenamefont {Deutsch},
  \citenamefont {H{\"a}nsch},\ and\ \citenamefont {Reichel}}]{hunger2010fiber}%
  \BibitemOpen
  \bibfield  {author} {\bibinfo {author} {\bibfnamefont {D.}~\bibnamefont
  {Hunger}}, \bibinfo {author} {\bibfnamefont {T.}~\bibnamefont {Steinmetz}},
  \bibinfo {author} {\bibfnamefont {Y.}~\bibnamefont {Colombe}}, \bibinfo
  {author} {\bibfnamefont {C.}~\bibnamefont {Deutsch}}, \bibinfo {author}
  {\bibfnamefont {T.~W.}\ \bibnamefont {H{\"a}nsch}},\ and\ \bibinfo {author}
  {\bibfnamefont {J.}~\bibnamefont {Reichel}},\ }\bibfield  {title} {\bibinfo
  {title} {A fiber fabry--perot cavity with high finesse},\ }\href@noop {}
  {\bibfield  {journal} {\bibinfo  {journal} {New Journal of Physics}\ }\textbf
  {\bibinfo {volume} {12}},\ \bibinfo {pages} {065038} (\bibinfo {year}
  {2010})}\BibitemShut {NoStop}%
\bibitem [{\citenamefont {Jiang}\ \emph {et~al.}(2019)\citenamefont {Jiang},
  \citenamefont {Patel}, \citenamefont {Mayor}, \citenamefont {McKenna},
  \citenamefont {Arrangoiz-Arriola}, \citenamefont {Sarabalis}, \citenamefont
  {Witmer}, \citenamefont {van Laer},\ and\ \citenamefont
  {Safavi-Naeini}}]{jiang_lithium_2019}%
  \BibitemOpen
  \bibfield  {author} {\bibinfo {author} {\bibfnamefont {W.}~\bibnamefont
  {Jiang}}, \bibinfo {author} {\bibfnamefont {R.~N.}\ \bibnamefont {Patel}},
  \bibinfo {author} {\bibfnamefont {F.~M.}\ \bibnamefont {Mayor}}, \bibinfo
  {author} {\bibfnamefont {T.~P.}\ \bibnamefont {McKenna}}, \bibinfo {author}
  {\bibfnamefont {P.}~\bibnamefont {Arrangoiz-Arriola}}, \bibinfo {author}
  {\bibfnamefont {C.~J.}\ \bibnamefont {Sarabalis}}, \bibinfo {author}
  {\bibfnamefont {J.~D.}\ \bibnamefont {Witmer}}, \bibinfo {author}
  {\bibfnamefont {R.}~\bibnamefont {van Laer}},\ and\ \bibinfo {author}
  {\bibfnamefont {A.~H.}\ \bibnamefont {Safavi-Naeini}},\ }\bibfield  {title}
  {\bibinfo {title} {{Lithium niobate piezo-optomechanical crystals}},\
  }\href@noop {} {\bibfield  {journal} {\bibinfo  {journal} {Optica}\ }\textbf
  {\bibinfo {volume} {6}},\ \bibinfo {pages} {845} (\bibinfo {year}
  {2019})}\BibitemShut {NoStop}%
\bibitem [{\citenamefont {Ramp}\ \emph {et~al.}(2019)\citenamefont {Ramp},
  \citenamefont {Hauer}, \citenamefont {Balram}, \citenamefont {Clark},
  \citenamefont {Srinivasan},\ and\ \citenamefont
  {Davis}}]{ramp_elimination_2018}%
  \BibitemOpen
  \bibfield  {author} {\bibinfo {author} {\bibfnamefont {H.}~\bibnamefont
  {Ramp}}, \bibinfo {author} {\bibfnamefont {B.~D.}\ \bibnamefont {Hauer}},
  \bibinfo {author} {\bibfnamefont {K.~C.}\ \bibnamefont {Balram}}, \bibinfo
  {author} {\bibfnamefont {T.~J.}\ \bibnamefont {Clark}}, \bibinfo {author}
  {\bibfnamefont {K.}~\bibnamefont {Srinivasan}},\ and\ \bibinfo {author}
  {\bibfnamefont {J.~P.}\ \bibnamefont {Davis}},\ }\bibfield  {title} {\bibinfo
  {title} {Elimination of thermomechanical noise in piezoelectric
  optomechanical crystals},\ }\href
  {https://doi.org/10.1103/PhysRevLett.123.093603} {\bibfield  {journal}
  {\bibinfo  {journal} {Phys. Rev. Lett.}\ }\textbf {\bibinfo {volume} {123}},\
  \bibinfo {pages} {093603} (\bibinfo {year} {2019})}\BibitemShut {NoStop}%
\bibitem [{\citenamefont {Datta}(1986)}]{datta1986surface}%
  \BibitemOpen
  \bibfield  {author} {\bibinfo {author} {\bibfnamefont {S.}~\bibnamefont
  {Datta}},\ }\href@noop {} {\emph {\bibinfo {title} {Surface acoustic wave
  devices}}}\ (\bibinfo  {publisher} {Prentice Hall},\ \bibinfo {year}
  {1986})\BibitemShut {NoStop}%
\bibitem [{\citenamefont {Siddiqui}\ \emph {et~al.}(2018)\citenamefont
  {Siddiqui}, \citenamefont {Olsson},\ and\ \citenamefont
  {Eichenfield}}]{siddiqui2018lamb}%
  \BibitemOpen
  \bibfield  {author} {\bibinfo {author} {\bibfnamefont {A.}~\bibnamefont
  {Siddiqui}}, \bibinfo {author} {\bibfnamefont {R.~H.}\ \bibnamefont
  {Olsson}},\ and\ \bibinfo {author} {\bibfnamefont {M.}~\bibnamefont
  {Eichenfield}},\ }\bibfield  {title} {\bibinfo {title} {Lamb wave focusing
  transducer for efficient coupling to wavelength-scale structures in thin
  piezoelectric films},\ }\href@noop {} {\bibfield  {journal} {\bibinfo
  {journal} {Journal of Microelectromechanical Systems}\ }\textbf {\bibinfo
  {volume} {27}},\ \bibinfo {pages} {1054} (\bibinfo {year}
  {2018})}\BibitemShut {NoStop}%
\bibitem [{\citenamefont {de~Lima~Jr}\ \emph {et~al.}(2003)\citenamefont
  {de~Lima~Jr}, \citenamefont {Alsina}, \citenamefont {Seidel},\ and\
  \citenamefont {Santos}}]{de2003focusing}%
  \BibitemOpen
  \bibfield  {author} {\bibinfo {author} {\bibfnamefont {M.}~\bibnamefont
  {de~Lima~Jr}}, \bibinfo {author} {\bibfnamefont {F.}~\bibnamefont {Alsina}},
  \bibinfo {author} {\bibfnamefont {W.}~\bibnamefont {Seidel}},\ and\ \bibinfo
  {author} {\bibfnamefont {P.}~\bibnamefont {Santos}},\ }\bibfield  {title}
  {\bibinfo {title} {Focusing of surface-acoustic-wave fields on (100) gaas
  surfaces},\ }\href@noop {} {\bibfield  {journal} {\bibinfo  {journal}
  {Journal of applied physics}\ }\textbf {\bibinfo {volume} {94}},\ \bibinfo
  {pages} {7848} (\bibinfo {year} {2003})}\BibitemShut {NoStop}%
\bibitem [{\citenamefont {Hopcroft}\ \emph {et~al.}(2010)\citenamefont
  {Hopcroft}, \citenamefont {Nix},\ and\ \citenamefont
  {Kenny}}]{hopcroft2010young}%
  \BibitemOpen
  \bibfield  {author} {\bibinfo {author} {\bibfnamefont {M.~A.}\ \bibnamefont
  {Hopcroft}}, \bibinfo {author} {\bibfnamefont {W.~D.}\ \bibnamefont {Nix}},\
  and\ \bibinfo {author} {\bibfnamefont {T.~W.}\ \bibnamefont {Kenny}},\
  }\bibfield  {title} {\bibinfo {title} {What is the young's modulus of
  silicon?},\ }\href@noop {} {\bibfield  {journal} {\bibinfo  {journal}
  {Journal of microelectromechanical systems}\ }\textbf {\bibinfo {volume}
  {19}},\ \bibinfo {pages} {229} (\bibinfo {year} {2010})}\BibitemShut
  {NoStop}%
\bibitem [{\citenamefont {MacCabe}\ \emph
  {et~al.}(2020{\natexlab{b}})\citenamefont {MacCabe}, \citenamefont {Ren},
  \citenamefont {Luo}, \citenamefont {Cohen}, \citenamefont {Zhou},
  \citenamefont {Sipahigil}, \citenamefont {Mirhosseini},\ and\ \citenamefont
  {Painter}}]{macCabe_phononic_2019}%
  \BibitemOpen
  \bibfield  {author} {\bibinfo {author} {\bibfnamefont {G.~S.}\ \bibnamefont
  {MacCabe}}, \bibinfo {author} {\bibfnamefont {H.}~\bibnamefont {Ren}},
  \bibinfo {author} {\bibfnamefont {J.}~\bibnamefont {Luo}}, \bibinfo {author}
  {\bibfnamefont {J.~D.}\ \bibnamefont {Cohen}}, \bibinfo {author}
  {\bibfnamefont {H.}~\bibnamefont {Zhou}}, \bibinfo {author} {\bibfnamefont
  {A.}~\bibnamefont {Sipahigil}}, \bibinfo {author} {\bibfnamefont
  {M.}~\bibnamefont {Mirhosseini}},\ and\ \bibinfo {author} {\bibfnamefont
  {O.}~\bibnamefont {Painter}},\ }\bibfield  {title} {\bibinfo {title}
  {Nano-acoustic resonator with ultralong phonon lifetime},\ }\href
  {https://doi.org/10.1126/science.abc7312} {\bibfield  {journal} {\bibinfo
  {journal} {Science}\ }\textbf {\bibinfo {volume} {370}},\ \bibinfo {pages}
  {840} (\bibinfo {year} {2020}{\natexlab{b}})}\BibitemShut {NoStop}%
\bibitem [{\citenamefont {Olsson~III}\ \emph {et~al.}(2014)\citenamefont
  {Olsson~III}, \citenamefont {Hattar}, \citenamefont {Homeijer}, \citenamefont
  {Wiwi}, \citenamefont {Eichenfield}, \citenamefont {Branch}, \citenamefont
  {Baker}, \citenamefont {Nguyen}, \citenamefont {Clark}, \citenamefont {Bauer}
  \emph {et~al.}}]{olsson2014high}%
  \BibitemOpen
  \bibfield  {author} {\bibinfo {author} {\bibfnamefont {R.~H.}\ \bibnamefont
  {Olsson~III}}, \bibinfo {author} {\bibfnamefont {K.}~\bibnamefont {Hattar}},
  \bibinfo {author} {\bibfnamefont {S.~J.}\ \bibnamefont {Homeijer}}, \bibinfo
  {author} {\bibfnamefont {M.}~\bibnamefont {Wiwi}}, \bibinfo {author}
  {\bibfnamefont {M.}~\bibnamefont {Eichenfield}}, \bibinfo {author}
  {\bibfnamefont {D.~W.}\ \bibnamefont {Branch}}, \bibinfo {author}
  {\bibfnamefont {M.~S.}\ \bibnamefont {Baker}}, \bibinfo {author}
  {\bibfnamefont {J.}~\bibnamefont {Nguyen}}, \bibinfo {author} {\bibfnamefont
  {B.}~\bibnamefont {Clark}}, \bibinfo {author} {\bibfnamefont
  {T.}~\bibnamefont {Bauer}}, \emph {et~al.},\ }\bibfield  {title} {\bibinfo
  {title} {A high electromechanical coupling coefficient sh0 lamb wave lithium
  niobate micromechanical resonator and a method for fabrication},\ }\href@noop
  {} {\bibfield  {journal} {\bibinfo  {journal} {Sensors and Actuators A:
  Physical}\ }\textbf {\bibinfo {volume} {209}},\ \bibinfo {pages} {183}
  (\bibinfo {year} {2014})}\BibitemShut {NoStop}%
\bibitem [{\citenamefont {Ju}\ \emph {et~al.}(2000)\citenamefont {Ju},
  \citenamefont {Blair},\ and\ \citenamefont {Zhao}}]{ju2000detection}%
  \BibitemOpen
  \bibfield  {author} {\bibinfo {author} {\bibfnamefont {L.}~\bibnamefont
  {Ju}}, \bibinfo {author} {\bibfnamefont {D.}~\bibnamefont {Blair}},\ and\
  \bibinfo {author} {\bibfnamefont {C.}~\bibnamefont {Zhao}},\ }\bibfield
  {title} {\bibinfo {title} {Detection of gravitational waves},\ }\href@noop {}
  {\bibfield  {journal} {\bibinfo  {journal} {Reports on Progress in Physics}\
  }\textbf {\bibinfo {volume} {63}},\ \bibinfo {pages} {1317} (\bibinfo {year}
  {2000})}\BibitemShut {NoStop}%
\bibitem [{\citenamefont {Tadesse}\ \emph {et~al.}(2015)\citenamefont
  {Tadesse}, \citenamefont {Li}, \citenamefont {Liu},\ and\ \citenamefont
  {Li}}]{tadesse_acousto-optic_2015}%
  \BibitemOpen
  \bibfield  {author} {\bibinfo {author} {\bibfnamefont {S.~A.}\ \bibnamefont
  {Tadesse}}, \bibinfo {author} {\bibfnamefont {H.}~\bibnamefont {Li}},
  \bibinfo {author} {\bibfnamefont {Q.}~\bibnamefont {Liu}},\ and\ \bibinfo
  {author} {\bibfnamefont {M.}~\bibnamefont {Li}},\ }\bibfield  {title}
  {\bibinfo {title} {Acousto-optic modulation of a photonic crystal nanocavity
  with {Lamb} waves in microwave {K} band},\ }\href@noop {} {\bibfield
  {journal} {\bibinfo  {journal} {Appl. Phys. Lett.}\ }\textbf {\bibinfo
  {volume} {107}},\ \bibinfo {pages} {201113} (\bibinfo {year}
  {2015})}\BibitemShut {NoStop}%
\bibitem [{\citenamefont {Shao}\ \emph {et~al.}(2019)\citenamefont {Shao},
  \citenamefont {Yu}, \citenamefont {Maity}, \citenamefont {Sinclair},
  \citenamefont {Zheng}, \citenamefont {Chia}, \citenamefont {Shams-Ansari},
  \citenamefont {Wang}, \citenamefont {Zhang}, \citenamefont {Lai} \emph
  {et~al.}}]{shao2019microwave}%
  \BibitemOpen
  \bibfield  {author} {\bibinfo {author} {\bibfnamefont {L.}~\bibnamefont
  {Shao}}, \bibinfo {author} {\bibfnamefont {M.}~\bibnamefont {Yu}}, \bibinfo
  {author} {\bibfnamefont {S.}~\bibnamefont {Maity}}, \bibinfo {author}
  {\bibfnamefont {N.}~\bibnamefont {Sinclair}}, \bibinfo {author}
  {\bibfnamefont {L.}~\bibnamefont {Zheng}}, \bibinfo {author} {\bibfnamefont
  {C.}~\bibnamefont {Chia}}, \bibinfo {author} {\bibfnamefont {A.}~\bibnamefont
  {Shams-Ansari}}, \bibinfo {author} {\bibfnamefont {C.}~\bibnamefont {Wang}},
  \bibinfo {author} {\bibfnamefont {M.}~\bibnamefont {Zhang}}, \bibinfo
  {author} {\bibfnamefont {K.}~\bibnamefont {Lai}}, \emph {et~al.},\ }\bibfield
   {title} {\bibinfo {title} {Microwave-to-optical conversion using lithium
  niobate thin-film acoustic resonators},\ }\href@noop {} {\bibfield  {journal}
  {\bibinfo  {journal} {Optica}\ }\textbf {\bibinfo {volume} {6}},\ \bibinfo
  {pages} {1498} (\bibinfo {year} {2019})}\BibitemShut {NoStop}%
\bibitem [{\citenamefont {Fuhrmann}\ \emph {et~al.}(2011)\citenamefont
  {Fuhrmann}, \citenamefont {Thon}, \citenamefont {Kim}, \citenamefont
  {Bouwmeester}, \citenamefont {Petroff}, \citenamefont {Wixforth},\ and\
  \citenamefont {Krenner}}]{fuhrmann2011dynamic}%
  \BibitemOpen
  \bibfield  {author} {\bibinfo {author} {\bibfnamefont {D.~A.}\ \bibnamefont
  {Fuhrmann}}, \bibinfo {author} {\bibfnamefont {S.~M.}\ \bibnamefont {Thon}},
  \bibinfo {author} {\bibfnamefont {H.}~\bibnamefont {Kim}}, \bibinfo {author}
  {\bibfnamefont {D.}~\bibnamefont {Bouwmeester}}, \bibinfo {author}
  {\bibfnamefont {P.~M.}\ \bibnamefont {Petroff}}, \bibinfo {author}
  {\bibfnamefont {A.}~\bibnamefont {Wixforth}},\ and\ \bibinfo {author}
  {\bibfnamefont {H.~J.}\ \bibnamefont {Krenner}},\ }\bibfield  {title}
  {\bibinfo {title} {Dynamic modulation of photonic crystal nanocavities using
  gigahertz acoustic phonons},\ }\href@noop {} {\bibfield  {journal} {\bibinfo
  {journal} {Nature Photonics}\ }\textbf {\bibinfo {volume} {5}},\ \bibinfo
  {pages} {605} (\bibinfo {year} {2011})}\BibitemShut {NoStop}%
\bibitem [{\citenamefont {Jiang}\ and\ \citenamefont
  {Balram}(2020)}]{jiang2020suspended}%
  \BibitemOpen
  \bibfield  {author} {\bibinfo {author} {\bibfnamefont {P.}~\bibnamefont
  {Jiang}}\ and\ \bibinfo {author} {\bibfnamefont {K.~C.}\ \bibnamefont
  {Balram}},\ }\bibfield  {title} {\bibinfo {title} {Suspended gallium arsenide
  platform for building large scale photonic integrated circuits: passive
  devices},\ }\href@noop {} {\bibfield  {journal} {\bibinfo  {journal} {Optics
  express}\ }\textbf {\bibinfo {volume} {28}},\ \bibinfo {pages} {12262}
  (\bibinfo {year} {2020})}\BibitemShut {NoStop}%
\bibitem [{\citenamefont {Forsch}\ \emph
  {et~al.}(2020{\natexlab{b}})\citenamefont {Forsch}, \citenamefont {Stockill},
  \citenamefont {Wallucks}, \citenamefont {Marinkovi{\'c}}, \citenamefont
  {G{\"a}rtner}, \citenamefont {Norte}, \citenamefont {van Otten},
  \citenamefont {Fiore}, \citenamefont {Srinivasan},\ and\ \citenamefont
  {Groeblacher}}]{forsch2020microwave}%
  \BibitemOpen
  \bibfield  {author} {\bibinfo {author} {\bibfnamefont {M.}~\bibnamefont
  {Forsch}}, \bibinfo {author} {\bibfnamefont {R.}~\bibnamefont {Stockill}},
  \bibinfo {author} {\bibfnamefont {A.}~\bibnamefont {Wallucks}}, \bibinfo
  {author} {\bibfnamefont {I.}~\bibnamefont {Marinkovi{\'c}}}, \bibinfo
  {author} {\bibfnamefont {C.}~\bibnamefont {G{\"a}rtner}}, \bibinfo {author}
  {\bibfnamefont {R.~A.}\ \bibnamefont {Norte}}, \bibinfo {author}
  {\bibfnamefont {F.}~\bibnamefont {van Otten}}, \bibinfo {author}
  {\bibfnamefont {A.}~\bibnamefont {Fiore}}, \bibinfo {author} {\bibfnamefont
  {K.}~\bibnamefont {Srinivasan}},\ and\ \bibinfo {author} {\bibfnamefont
  {S.}~\bibnamefont {Groeblacher}},\ }\bibfield  {title} {\bibinfo {title}
  {Microwave-to-optics conversion using a mechanical oscillator in its quantum
  ground state},\ }\href@noop {} {\bibfield  {journal} {\bibinfo  {journal}
  {Nature Physics}\ }\textbf {\bibinfo {volume} {16}},\ \bibinfo {pages} {69}
  (\bibinfo {year} {2020}{\natexlab{b}})}\BibitemShut {NoStop}%
\bibitem [{\citenamefont {Najer}\ \emph {et~al.}(2019)\citenamefont {Najer},
  \citenamefont {S{\"o}llner}, \citenamefont {Sekatski}, \citenamefont
  {Dolique}, \citenamefont {L{\"o}bl}, \citenamefont {Riedel}, \citenamefont
  {Schott}, \citenamefont {Starosielec}, \citenamefont {Valentin},
  \citenamefont {Wieck} \emph {et~al.}}]{najer2019gated}%
  \BibitemOpen
  \bibfield  {author} {\bibinfo {author} {\bibfnamefont {D.}~\bibnamefont
  {Najer}}, \bibinfo {author} {\bibfnamefont {I.}~\bibnamefont {S{\"o}llner}},
  \bibinfo {author} {\bibfnamefont {P.}~\bibnamefont {Sekatski}}, \bibinfo
  {author} {\bibfnamefont {V.}~\bibnamefont {Dolique}}, \bibinfo {author}
  {\bibfnamefont {M.~C.}\ \bibnamefont {L{\"o}bl}}, \bibinfo {author}
  {\bibfnamefont {D.}~\bibnamefont {Riedel}}, \bibinfo {author} {\bibfnamefont
  {R.}~\bibnamefont {Schott}}, \bibinfo {author} {\bibfnamefont
  {S.}~\bibnamefont {Starosielec}}, \bibinfo {author} {\bibfnamefont {S.~R.}\
  \bibnamefont {Valentin}}, \bibinfo {author} {\bibfnamefont {A.~D.}\
  \bibnamefont {Wieck}}, \emph {et~al.},\ }\bibfield  {title} {\bibinfo {title}
  {A gated quantum dot strongly coupled to an optical microcavity},\
  }\href@noop {} {\bibfield  {journal} {\bibinfo  {journal} {Nature}\ }\textbf
  {\bibinfo {volume} {575}},\ \bibinfo {pages} {622} (\bibinfo {year}
  {2019})}\BibitemShut {NoStop}%
\bibitem [{\citenamefont {Van~Laer}\ \emph {et~al.}(2020)\citenamefont
  {Van~Laer}, \citenamefont {Jiang}, \citenamefont {Patel}, \citenamefont
  {Sarabalis}, \citenamefont {Cleland}, \citenamefont {McKenna}, \citenamefont
  {Wollack}, \citenamefont {Arrangoiz-Arriola}, \citenamefont {Witmer},\ and\
  \citenamefont {Safavi-Naeini}}]{van2020piezo}%
  \BibitemOpen
  \bibfield  {author} {\bibinfo {author} {\bibfnamefont {R.}~\bibnamefont
  {Van~Laer}}, \bibinfo {author} {\bibfnamefont {W.}~\bibnamefont {Jiang}},
  \bibinfo {author} {\bibfnamefont {R.~N.}\ \bibnamefont {Patel}}, \bibinfo
  {author} {\bibfnamefont {C.~J.}\ \bibnamefont {Sarabalis}}, \bibinfo {author}
  {\bibfnamefont {A.}~\bibnamefont {Cleland}}, \bibinfo {author} {\bibfnamefont
  {T.~P.}\ \bibnamefont {McKenna}}, \bibinfo {author} {\bibfnamefont {E.~A.}\
  \bibnamefont {Wollack}}, \bibinfo {author} {\bibfnamefont {P.}~\bibnamefont
  {Arrangoiz-Arriola}}, \bibinfo {author} {\bibfnamefont {J.~D.}\ \bibnamefont
  {Witmer}},\ and\ \bibinfo {author} {\bibfnamefont {A.~H.}\ \bibnamefont
  {Safavi-Naeini}},\ }\bibfield  {title} {\bibinfo {title} {Piezo-optomechanics
  in lithium niobate on silicon-on-insulator for microwave-to-optics
  transduction},\ }in\ \href@noop {} {\emph {\bibinfo {booktitle} {CLEO:
  Science and Innovations}}}\ (\bibinfo {organization} {Optical Society of
  America},\ \bibinfo {year} {2020})\ pp.\ \bibinfo {pages}
  {STu4J--2}\BibitemShut {NoStop}%
\bibitem [{\citenamefont {Marinkovic}\ \emph {et~al.}(2021)\citenamefont
  {Marinkovic}, \citenamefont {Drimmer}, \citenamefont {Hensen},\ and\
  \citenamefont {Groeblacher}}]{marinkovic2021hybrid}%
  \BibitemOpen
  \bibfield  {author} {\bibinfo {author} {\bibfnamefont {I.}~\bibnamefont
  {Marinkovic}}, \bibinfo {author} {\bibfnamefont {M.}~\bibnamefont {Drimmer}},
  \bibinfo {author} {\bibfnamefont {B.}~\bibnamefont {Hensen}},\ and\ \bibinfo
  {author} {\bibfnamefont {S.}~\bibnamefont {Groeblacher}},\ }\bibfield
  {title} {\bibinfo {title} {Hybrid integration of silicon photonic devices on
  lithium niobate for optomechanical wavelength conversion},\ }\href@noop {}
  {\bibfield  {journal} {\bibinfo  {journal} {Nano letters}\ }\textbf {\bibinfo
  {volume} {21}},\ \bibinfo {pages} {529} (\bibinfo {year} {2021})}\BibitemShut
  {NoStop}%
\bibitem [{Note1()}]{Note1}%
  \BibitemOpen
  \bibinfo {note} {Certain commercial products or company names are identified
  here to describe our study adequately. Such identification is not intended to
  imply recommendation or endorsement by the National Institute of Standards
  and Technology, nor is it intended to imply that the products or names
  identified are necessarily the best available for the purpose.}\BibitemShut
  {Stop}%
\bibitem [{\citenamefont {Lecocq}\ \emph {et~al.}(2016)\citenamefont {Lecocq},
  \citenamefont {Clark}, \citenamefont {Simmonds}, \citenamefont {Aumentado},\
  and\ \citenamefont {Teufel}}]{lecocq_mechanically_2016}%
  \BibitemOpen
  \bibfield  {author} {\bibinfo {author} {\bibfnamefont {F.}~\bibnamefont
  {Lecocq}}, \bibinfo {author} {\bibfnamefont {J.}~\bibnamefont {Clark}},
  \bibinfo {author} {\bibfnamefont {R.}~\bibnamefont {Simmonds}}, \bibinfo
  {author} {\bibfnamefont {J.}~\bibnamefont {Aumentado}},\ and\ \bibinfo
  {author} {\bibfnamefont {J.}~\bibnamefont {Teufel}},\ }\bibfield  {title}
  {\bibinfo {title} {Mechanically {Mediated} {Microwave} {Frequency}
  {Conversion} in the {Quantum} {Regime}},\ }\href@noop {} {\bibfield
  {journal} {\bibinfo  {journal} {Physical Review Letters}\ }\textbf {\bibinfo
  {volume} {116}},\ \bibinfo {pages} {043601} (\bibinfo {year}
  {2016})}\BibitemShut {NoStop}%
\end{thebibliography}%

\end{document}